\documentclass[12pt]{article}
\usepackage{amssymb}
\usepackage{amsfonts}
\usepackage{amsmath}

\renewcommand{\baselinestretch}{1.5}
\newtheorem{theorem}{Theorem}

\newtheorem{claim}{Claim}

\newtheorem{corollary}{Corollary}

\newtheorem{lemma}{Lemma}

\newtheorem{proposition}{Proposition}
\newtheorem{remark}{Remark}

\begin{document}

\title{Strategic Analysis of Fair Rank-Minimizing Mechanisms with Agent
Refusal Option\thanks{%
This work was supported by JSPS KAKENHI Grant Numbers 22K01402 and 24K04932.}%
}
\author{Yasunori Okumura\thanks{%
Department of Logistics and Information Engineering, TUMSAT, 2-1-6,
Etchujima, Koto-ku, Tokyo, 135-8533 Japan. Phone:+81-3-5245-7300.
Fax:+81-3-5245-7300. E-mail: okuyasu@gs.econ.keio.ac.jp}}

\maketitle

\begin{center}
\textbf{Abstract}
\end{center}

This paper investigates the strategic implications of the uniform
rank-minimizing mechanism (URM), an assignment rule that selects uniformly
from the set of deterministic assignments minimizing the sum of agents'
reported ranks. We focus on settings in which agents may refuse their
assignment and instead receive an outside option. Without the refusal
option, we show that truth-telling is not strictly dominated under any fair
rank-minimizing mechanism; that is, one satisfying equal treatment of
equals. However, introducing the refusal option significantly changes
strategic incentives: specific manipulations, called outside option demotion
strategies, dominate truth-telling under the URM. Moreover, such
manipulations can lead to inefficient outcomes, as desirable objects may be
refused by misreporting agents and consequently remain unassigned. To
address this issue, we propose a modification of the URM that restores
undominated truth-telling, although it introduces incentives to underreport
acceptable objects. Our results highlight a fundamental trade-off in the
design of fair rank-minimizing mechanisms when agents can refuse their
assignments.


\newpage

\section{Introduction}

Many assignment problems such as school choice or job assignment require
matching agents with heterogeneous objects based on ordinal preferences. In
such contexts, policymakers often seek to minimize the sum (or average) of
the ranks of assigned objects. For instance, Featherstone (2020) observes
that Teach For America explicitly considers rank distribution in its
assignment decisions. However, the authority can optimize only the ranks
reported by individuals, which may differ from their true preference ranks.
These considerations motivate the study of rank-minimizing mechanisms, which
determine assignments to minimize the total reported ranks.

A well-known example is the uniform rank-minimizing mechanism (hereafter,
URM), which selects uniformly at random from the set of deterministic
assignments that minimize the total rank. The URM satisfies equal treatment
of equals (Bogomolnaia and Moulin, 2001), a fundamental fairness notion
requiring agents who report identical preferences to receive identical
(ex-ante) treatment. Mechanisms that are both rank-minimizing and satisfy
this fairness condition are referred to as fair rank-minimizing mechanisms,
with the URM being their most representative example.

While it is known that rank-minimizing mechanisms are not strategy-proof
(Featherstone, 2020), we mainly study whether truth-telling is a dominated
strategy. We first consider the benchmark setting without any refusal
options, where each agent must accept the assigned object. In this case, we
show that truth-telling is not a dominated strategy.

Our main analysis concerns the case in which agents can refuse their
assignment and instead receive an outside option.\footnote{%
Feigenbaum et al. (2020) and Afacan (2022) state that in some real school
choice markets, such as public school markets in Boston, New York City, and
Turkey, students actually refuse their assignments. For example, about 10\%
of the students withdrew from the New York City public high school placement.%
} In this setting, we show that a class of strategies called outside option
demotion strategies, which place the outside option at the bottom of the
reported ranking while preserving the order over acceptable objects, can
strictly dominate truth-telling under the URM.

The intuition is as follows: for some preference profiles of other agents,
an agent who adopts such a strategy may succeed in avoiding the outside
option and instead receive a truly acceptable object, because assigning the
outside option to that agent would substantially increase the total rank. In
the standard setting, this manipulation would be risky, as the agent might
be assigned an unacceptable object. However, with the refusal option in
place, the agent can simply reject any such assignment and receive the
outside option instead. Thus, the strategies become effectively risk-free.

We further show that such strategies may lead to inefficient outcomes: some
objects that are desired by agents may be left unassigned if misreporting
agents refuse them. In addition, we provide a concrete example in which an
outside option demotion strategy forms part of a pure-strategy Nash
equilibrium, reinforcing the strategic vulnerability of the URM with the
refusal option.

Finally, we propose a modified version of the URM that prevents
truth-telling from being strictly dominated and is fair. However, this
mechanism introduces new incentives: agents may benefit by promoting the
outside option. Our results thus reveal a fundamental trade-off in the
design of fair rank-minimizing assignment mechanisms in the presence of
refusal options.

\subsection*{Related Literature}

There is an extensive literature on fair assignment in probabilistic
assignment problems. Prominent studies include Bogomolnaia and Moulin
(2001), Budish et al. (2013), and Nesterov (2017), who explore fairness
notions such as weak envy-freeness and the equal division lower bound.
However, Feizi (2024) shows that these fairness notions are incompatible
with rank-minimization. In contrast, ETE, which we adopt in this study, is
compatible with rank-minimizing mechanisms and thus suitable for our
analysis.

The efficiency of rank-minimizing mechanisms is recently also well studied.
Featherstone (2020) and Feizi (2024) show that rank-minimizing mechanisms
satisfy stronger efficiency properties than commonly used ones, including
ordinal efficiency, originally introduced by Bogomolnaia and Moulin (2001).%
\footnote{%
Although Featherstone (2020) introduce the concept of rank-minimizing
assignments in the economic literature, a broader body of work exists in
mathematics and operations research. See Krokhmal and Pardalos (2009) for a
comprehensive survey.} Ortega and Klein (2023) compare outcomes under
rank-minimizing mechanisms with those from the deferred acceptance and top
trading cycles mechanisms, showing that the former improve on the latter two
in several dimensions of efficiency.

Regarding strategic behavior, much of the existing literature focuses on
settings without the refusal option. In this context, Nikzad (2022)
establishes Bayesian incentive compatibility for the URM in uniform markets,
and Troyan (2024) shows that the URM is not obviously manipulable, in the
sense of Troyan and Morrill (2020). Our result in the no-refusal setting is
related but logically independent: we show that under any fair
rank-minimizing mechanism, truth-telling is not a dominated strategy.

On the other hand, several studies have pointed out strategic issues
associated with rank-minimizing mechanisms even without the refusal option.
Tasnim et al. (2024) propose strategies for improving individual profit
under rank-minimizing mechanisms and show, using both simulated markets and
the real matching market in Amsterdam, that applying these strategies can
potentially reduce overall matching performance. Bando et al. (2025) also
introduce a computationally efficient algorithm for finding profitable
manipulations when agents are pessimistic. Our study complements them by
theoretically identifying strategic dominance relations in settings where
agents can refuse their assignments.

Featherstone (2020) considers settings where agents can refuse their
assignment. His analysis assumes that agents possess limited
information---specifically, they are unaware of other agents' preferences
and of the capacities of all objects except their own outside option. Our
model departs from this assumption: we suppose that agents know the
capacities of all objects, as may be the case in a school choice context
where this information is publicly available. Moreover, we abstract from
assumptions about agents' beliefs or rationality regarding others, focusing
mainly on strategic dominance.

Featherstone (2020) shows that when information is limited as above and
agents have the refusal option, revealing the full extension, which is an
instance of an ODS, is weakly better than any alternative. His results imply
that rank-minimizing mechanisms are vulnerable to manipulation when outside
options are valued to some extent. Our findings show that this vulnerability
persists even in the absence of information limitation. Moreover, we show
that the full extension may itself be strictly dominated by another ODS when
agents are informed about object capacities. This highlights a new dimension
of strategic risk that does not arise under Featherstone's setting.

\section{Model}

We consider the model discussed by Kojima and Manea (2010) and Featherstone
(2020). Let $\mathcal{A}$ and $\mathcal{O}$ be finite sets of agents and
object types respectively. We assume $\left\vert \mathcal{A}\right\vert \geq
3$ and $\left\vert \mathcal{O}\right\vert \geq 3$. Each agent must be
allocated one copy of an object whose type is an element of a set denoted by 
$\mathcal{O}$. Let $q_{o}\geq 1$ be the number of copies (the capacity) of
an object type $o\in \mathcal{O}$. We assume that there is a \textbf{null
object} type denoted by $\varnothing \in \mathcal{O}$ satisfying $%
q_{\varnothing }\geq \left\vert \mathcal{A}\right\vert $. That is, the null
object type can be assigned to all agents simultaneously. The null object
type $\varnothing \in \mathcal{O}$ can be interpreted as the first-best
alternative object of an agent outside of this market. In Section 4, we
consider the situation where the null object type has the other
characteristic. To avoid trivial cases, we assume $q_{o}\in \left[
1,\left\vert \mathcal{A}\right\vert \right) $ for all $o\in \mathcal{%
O\setminus }\left\{ \varnothing \right\} $.

Let $\mathcal{P}$ be the set of all possible strict rankings of the elements
in $\mathcal{O}$. Let $p_{a}=\left( p_{a}^{1},\cdots ,p_{a}^{\left\vert 
\mathcal{O}\right\vert }\right) \in \mathcal{P}$ be a \textbf{preference
order} revealed by $a$, where $p_{a}^{k}=o$ means that type $o$ object is
ranked $k$th-best one under $p_{a}$. For notational convenience, we let $R:%
\mathcal{O\times P}\rightarrow \left\{ 1,\cdots ,\left\vert \mathcal{O}%
\right\vert \right\} $ be a rank function; that is, $p_{a}^{k}=o$ if and
only if $R\left( o,p_{a}\right) =k$. Moreover, if $R\left( o,p_{a}\right)
<R\left( \varnothing ,p_{a}\right) $ (resp. $R\left( o,p_{a}\right) >R\left(
\varnothing ,p_{a}\right) $), then $o$ is said to be \textbf{acceptable}
(resp. \textbf{unacceptable}) for $p_{a}$. Thus, $R\left( \varnothing
,p_{a}\right) -1$ represents the number of acceptable object types for $%
p_{a} $. Let $r_{a}\in \mathcal{P}$ be the \textbf{true preference order} of
agent $a$. Especially, if $R\left( o,r_{a}\right) <R\left( \varnothing
,r_{a}\right) $, then $o$ is said to be \textbf{truly} \textbf{acceptable}
for $a$.

An assignment is represented as lotteries over types. Let $x_{ao}$ represent
the probability that agent $a$ assigns (a copy of) type $o$ object.
Moreover, let $x$ be a matrix whose $\left( a,o\right) $ entry is equal to $%
x_{ao}$. We say that a matrix $x$ is an \textbf{assignment} if 
\begin{equation*}
x_{ao}\in \left[ 0,1\right] \text{, }\sum\limits_{o^{\prime }\in \mathcal{O}%
}x_{ao^{\prime }}=1\text{,\ and }\sum\limits_{a^{\prime }\in \mathcal{A}%
}x_{a^{\prime }o}\leq q_{o},
\end{equation*}%
for all $a\in \mathcal{A}$ and all $o\in \mathcal{O}$. Let $\mathcal{X}$ be
the set of all possible assignments. Specifically, $y\in \mathcal{X}$ is
said to be a \textbf{deterministic assignment} if $y_{ao}$ is either $0$ or $%
1$ for all $\left( a,o\right) \in \mathcal{A}\times \mathcal{O}$. Let $%
\mathcal{Y}\subseteq \mathcal{X}$ be a set of all possible deterministic
assignments.

The following result is due to Kojima and Manea (2010), which is a
generalization of the Birkhoff--von Neumann theorem.

\begin{remark}
Every assignment can be written as a convex combination of deterministic
assignments.
\end{remark}

We consider the preference of agent $a$ denoted by $p_{a}$ with regard to
two assignments $x$ and $x^{\prime }$. If 
\begin{equation}
\sum\limits_{o^{\prime }:\text{ }R\left( o^{\prime },p_{a}\right) \leq
k}x_{ao^{\prime }}\geq \sum\limits_{o^{\prime }:\text{ }R\left( o^{\prime
},p_{a}\right) \leq k}x_{ao^{\prime }}^{\prime }  \label{a}
\end{equation}%
for all $k=1,2,\cdots ,\left\vert \mathcal{O}\right\vert $, then agent $a$
with $p_{a}$ is said to \textbf{weakly prefer} (in the sense of first-order
stochastic dominance) $x$ to $x^{\prime }$. Moreover, $a$ is said to \textbf{%
strictly} \textbf{prefer} $x$ to $x^{\prime }$ if $a$ weakly prefers $x$ to $%
x^{\prime }$ and (\ref{a}) is satisfied with strict inequality for some $%
k=1,2,\cdots ,\left\vert \mathcal{O}\right\vert -1$.

Assignment $x$ is \textbf{wasteful }for $p$ if there are $o,o^{\prime }\in 
\mathcal{O}$ and $a\in \mathcal{A}$ such that $\sum\nolimits_{a^{\prime }\in 
\mathcal{A}}x_{a^{\prime }o}<q_{o}$, $x_{ao^{\prime }}>0$ and $R\left(
o,p_{a}\right) <R\left( o^{\prime },p_{a}\right) $. In words, an assignment
is considered wasteful for a given preference profile if there is at least
one agent who, despite the fact that a more desirable object is still
available, does not receive as much of it as they could, and instead
receives some amount of a less preferred object.

For an assignment $x\in \mathcal{X}$ and a preference profile $p=\left(
p_{a}\right) _{a\in \mathcal{A}}\in \mathcal{P}^{\left\vert \mathcal{A}%
\right\vert }$, let $RV\left( x,p\right) $ be such that 
\begin{equation*}
RV\left( x,p\right) =\sum\limits_{a\in \mathcal{A}}\sum\limits_{o\in 
\mathcal{O}}R\left( o,p_{a}\right) \times x_{ao},
\end{equation*}%
which is the sum of expected rank. An assignment $x^{\ast }\in \mathcal{X}$
is \textbf{rank-minimizing }for $p$\textbf{\ }if\textbf{\ }$RV\left( x^{\ast
},p\right) \leq RV\left( x,p\right) $ for all $x\in \mathcal{X}$. Let $%
\mathcal{X}^{\ast }\left( p\right) $ be the set of rank-minimizing
assignments for $p$. If $x^{\ast }\in \mathcal{X}$ is rank-minimizing for $p$%
, then $x$ is not wasteful\textbf{\ }for $p$.

In the subsequent section, we focus on the mechanisms whose input is $p\in 
\mathcal{P}^{\left\vert \mathcal{A}\right\vert }$ and output is a
rank-minimizing assignment for any $p\in \mathcal{P}^{\left\vert \mathcal{A}%
\right\vert }$. Before that we introduce some technical results.

\begin{lemma}
Fix any $x^{\ast }\in \mathcal{X\setminus Y}$. Let $y^{1},\cdots ,y^{I}\in 
\mathcal{Y}$ with $I\geq 2$ be such that%
\begin{equation*}
x^{\ast }=\sum\limits_{i=1}^{I}\sigma ^{i}y^{i}\in \mathcal{X}\text{ where }%
\sum\limits_{i=1}^{I}\sigma ^{i}=1
\end{equation*}%
and $\sigma ^{i}\in \left[ 0,1\right] $ for all $i=1,\cdots ,I$. If $x^{\ast
}$ is rank-minimizing,\textbf{\ }then $y^{i}$ is also rank-minimizing for
all $i=1,\cdots ,I$.
\end{lemma}

We provide the proofs of our results in the Appendix.

By Lemma 1, we immediately have the following result.

\begin{corollary}
There must exist at least one deterministic rank-minimizing assignment.
\end{corollary}

Nikzad (2022) and Troyan (2024) focus only on deterministic rank-minimizing
assignments; that is, they consider a deterministic assignment $y^{\ast }\in 
\mathcal{Y}$ such that $RV\left( y^{\ast },p\right) \leq RV\left( y,p\right) 
$ for all $y\in \mathcal{Y}$. However, by Corollary 1, such $y^{\ast }$ must
also be rank-minimizing in our definition; that is, $RV\left( y^{\ast
},p\right) \leq RV\left( x,p\right) $ for all $x\in \mathcal{X}$. Moreover,
this result allows us to show that the mechanism introduced later is
well-defined.

Finally, we introduce another technical result that is used to show our main
results introduced later. Let%
\begin{equation*}
\bar{k}\left( p_{a}\right) =\min \left\{ k\in \left\{ 1,\cdots ,\left\vert 
\mathcal{O}\right\vert \right\} \text{ }\left\vert \text{ }\sum\limits_{\bar{%
o}:R(\bar{o},p_{a})\leq k}q_{\bar{o}}\geq \left\vert \mathcal{A}\right\vert
\right. \right\} ,
\end{equation*}%
which is well-defined because $\sum\nolimits_{\bar{o}\in \mathcal{O}}q_{\bar{%
o}}>\left\vert \mathcal{A}\right\vert $. This means that for a given $%
p_{a}\in \mathcal{P}$, there are sufficient number of copies of object types
that ranked higher than or equal to $\bar{k}\left( p_{a}\right) $.
Trivially, if $p_{a}^{k}=\varnothing $, then $\bar{k}\left( p_{a}\right)
\leq k$. We have the following result.

\begin{lemma}
If an assignment $x\in \mathcal{X}$ is not wasteful\textbf{\ }for $p$, then $%
x_{ao}=0$ for all $o$ such that $R(o,p_{a})>\bar{k}\left( p_{a}\right) $.
\end{lemma}

This means that if an assignment is not wasteful for $p$, then $a$ must be
assigned to an object type that ranked higher than or equals to $\bar{k}%
\left( p_{a}\right) $. Typically, under a nonwasteful assignment, the
unacceptable object types in the preference ranking of agent $a$ are never
assigned to $a$, because $\varnothing $ is the object with a sufficient
number of copies.

\section{Rank-minimizing Mechanism}

In this section, we consider the following \textit{game}. All agents
simultaneously reveal a preference order in $\mathcal{P}$, which may or may
not be their true preference order $r_{a}\in \mathcal{P}$. Let $p_{a}\in 
\mathcal{P}$ be the preference order revealed by $a\in \mathcal{A}$ and $%
p=\left( p_{a}\right) _{a\in \mathcal{A}}\in \mathcal{P}^{\left\vert 
\mathcal{A}\right\vert }$. Then, based on $p$, an assignment is determined
via a mechanism.

We assume that all agents know the capacities of all object types $\left(
q_{o}\right) _{o\in \mathcal{O}}$ and the mechanism is employed to determine
the assignment, when they reveal a preference order. Note that in this
section, the null object type is just that with sufficient number of copies.

A mechanism employed in this game is represented by $f:\mathcal{P}%
^{\left\vert \mathcal{A}\right\vert }\rightarrow \mathcal{X}$. First, we
consider an efficiency property on mechanisms. We call $f$ a\textbf{\
rank-minimizing mechanism }if\textbf{\ }$f(p)$ is rank-minimizing for all $p$%
. In this study, we mainly focus on the following fair mechanism.

Let $\mathcal{Y}^{\ast }\left( p\right) \subseteq \mathcal{Y}$ be the set of
deterministic rank-minimizing assignments; that is, $y^{\ast }\in \mathcal{Y}%
^{\ast }\left( p\right) ,$ $RV\left( y^{\ast },p\right) \leq RV\left(
x,p\right) $ for any $x\in \mathcal{X}$. By Corollary 1, $\mathcal{Y}^{\ast
}\left( p\right) \neq \emptyset $. Let 
\begin{equation*}
f^{U}\left( p\right) =\sum\limits_{y^{\ast }\in \mathcal{Y}^{\ast }\left(
p\right) }\frac{y^{\ast }}{\left\vert \mathcal{Y}^{\ast }\left( p\right)
\right\vert }\text{ for all }p\in \mathcal{P}^{\left\vert \mathcal{A}%
\right\vert }
\end{equation*}%
be the \textbf{uniform} \textbf{rank-minimizing mechanism (URM)}. By (\ref{b}%
), 
\begin{equation*}
RV\left( f^{U}\left( p\right) ,p\right) =\sum\limits_{y^{\ast }\in \mathcal{Y%
}^{\ast }\left( p\right) }\frac{1}{\left\vert \mathcal{Y}^{\ast }\left(
p\right) \right\vert }RV\left( y^{\ast },p\right) .
\end{equation*}%
Thus, $RV\left( f^{U}\left( p\right) ,p\right) \leq RV\left( x,p\right) $
for any $x\in \mathcal{X}$; that is, $f^{U}\left( p\right) $ must be
rank-minimizing\textbf{\ }for $p$. Since Nikzad (2022), Ortega and Klein
(2023) and Troyan (2024) also discuss the URM, this is one of the most
common rank-minimizing mechanisms in the literature.

We show that the URM satisfies fundamental fairness properties. First, we
introduce a well-known one. A mechanism $f$ is said to satisfy \textbf{weak} 
\textbf{equal treatment of equals }(hereafter\textbf{\ weak ETE})\textbf{\ }%
if $p\in \mathcal{P}^{\left\vert \mathcal{A}\right\vert }$ satisfying $%
p_{a}=p_{a^{\prime }}$ implies $\left( f\left( p\right) \right) _{ao}=\left(
f\left( p\right) \right) _{a^{\prime }o}$ for all $o\in \mathcal{O}$. This
requires that two agents revealing the same preference must have the same
assignment in the ex-ante sense. This property is called just
\textquotedblleft equal treatment of equals\textquotedblright\ by many
previous studies such as Bogomolnaia and Moulin (2001) and Troyan (2024).

We can show that the URM satisfies a slightly stronger version. A mechanism $%
f$ is said to satisfy \textbf{equal treatment of equals (}hereafter \textbf{%
ETE) }if $p\in \mathcal{P}^{\left\vert \mathcal{A}\right\vert }$ satisfying $%
p_{a}^{k^{\prime }}=p_{a^{\prime }}^{k^{\prime }}$ for all $1,\cdots ,\bar{k}%
\left( p_{a}\right) $ implies $\left( f\left( p\right) \right) _{ao}=\left(
f\left( p\right) \right) _{a^{\prime }o}$ for all $o\in \mathcal{O}$.%
\footnote{%
Han (2024) also considers a similar property, called symmetricity, which is
slightly stronger than equal treatment of equals. Unlike in our model, they
take into account the priorities of objects and define agents as equal only
if they are equal with respect to the priorities of all objects.} By Lemma
2, as long as the assignment is not wasteful, any type that is ranked lower
than $\bar{k}\left( p_{a}\right) $th in the preference ranking revealed by $%
a $ will not be assigned to $a$, because there are sufficient copies of
types ranked higher than or equal to $\bar{k}\left( p_{a}\right) $th. Thus,
if two individuals reveal rankings that are identical for positions at or
above $\bar{k}\left( p_{a}\right) $th rank, those rankings are considered
essentially identical, and they should receive the same assignment in an
ex-ante sense.

We say that a mechanism satisfying ETE is a fair mechanism. We have the
following result.

\begin{proposition}
The URM satisfies ETE.
\end{proposition}

Therefore, the URM is a fair rank-minimizing one.

We discuss strategic problems of fair rank-minimizing mechanisms. We call
the strategy of revealing their true preference \textbf{truth-telling}.
Featherstone (2020, Proposition 10) shows that any rank-minimizing mechanism
does not satisfy strategy-proofness. Therefore, in this paper, we consider a
weaker strategic property.

For a mechanism $f$ and an agent $a$, (revealing) $p_{a}^{\prime }\in 
\mathcal{P}$ is said to be \textbf{weakly} (strategically\textbf{) dominated}
\textbf{by }(revealing) $p_{a}\in \mathcal{P}$ if $a$ weakly prefers $%
f\left( p_{a},p_{-a}\right) $ to $f\left( p_{a}^{\prime },p_{-a}\right) $
for all $p_{-a}\in \mathcal{P}^{\left\vert \mathcal{A}\right\vert -1}$. For $%
f$ and $a$, (revealing) $p_{a}^{\prime }$ is said to be (strictly\textbf{)
dominated} \textbf{by }(revealing) $p_{a}$ if $p_{a}^{\prime }$ is weakly
dominated by $p_{a}$ and there exists some $p_{-a}^{\prime }\in \mathcal{P}%
^{\left\vert \mathcal{A}\right\vert -1}$ such that $a$ strictly prefers $%
f\left( p_{a},p_{-a}^{\prime }\right) $ to $f\left( p_{a}^{\prime
},p_{-a}^{\prime }\right) $. Moreover, for $f$ and $a$, (revealing) $p_{a}$
is a (\textbf{weakly}) \textbf{dominated strategy} if there is some $%
p_{a}^{\prime }\in \mathcal{P}$ that (weakly) dominates $p_{a}$.

Moreover, we say that for $f$, $p\in \mathcal{P}^{\left\vert \mathcal{A}%
\right\vert }$ is a \textbf{Nash equilibrium} if there is no $a$ who
strictly prefers $f\left( p_{a}^{\prime },p_{-a}\right) $ to $f\left(
p_{a},p_{-a}\right) $ for some $p_{a}^{\prime }\in \mathcal{P}$. Therefore,
in this context, we define a Nash equilibrium as a profile of strategies
such that, for each agent, the resulting assignment is not first-order
stochastically dominated by the assignment that results if the agent
unilaterally deviated to any other strategy.

We mainly discuss whether truth-telling ($r_{a}$) is a dominated strategy.
Fix any $r_{a}$ and any $p_{a}\neq r_{a}$. Let $k$ be the smallest integer
such that $p_{ak}\neq r_{ak}$. By Lemma 2, if $k>\bar{k}\left( p_{a}\right) $%
, then $p_{a}$ and $r_{a}$ are essentially equivalent under a fair
rank-minimizing mechanism, because any result of a rank-minimizing mechanism
is not wasteful. Thus, we focus only on $r_{a}$ and $p_{a}$ satisfying $%
k\leq \bar{k}\left( p_{a}\right) $ defined above; that is,

\begin{equation*}
\sum\limits_{\bar{o}:R(\bar{o},p_{a})<k}q_{\bar{o}}\left( =\sum\limits_{\bar{%
o}:R(\bar{o},r_{a})<k}q_{\bar{o}}\right) <\left\vert \mathcal{A}\right\vert
\end{equation*}%
is satisfied for that $k$. In other word, we consider the case where the
number of object types that are more preferable than the $k-1$th best one in 
$p_{a}$ is insufficient to distribute to all agents. In this case, for some $%
p_{-a}\in \mathcal{P}^{\left\vert \mathcal{A}\right\vert -1}$ and some fair
rank-minimizing mechanism, $a$ is assigned to an object type that is less
preferable or equal to the $k$th-best one.

\begin{lemma}
Suppose that $f$ is a fair rank-minimizing mechanism. Let $p_{a}\neq r_{a}$,
and $k$ be the smallest integer such that $p_{a}^{k}\neq r_{a}^{k}$. If $%
k\leq \bar{k}\left( p_{a}\right) $, then there is $p_{-a}\in \mathcal{P}%
^{\left\vert \mathcal{A}\right\vert -1}$ such that agent $a$ (with $r_{a}$)
strictly prefers $f\left( r_{a},p_{-a}\right) $ to $f\left(
p_{a},p_{-a}\right) $.
\end{lemma}

By Lemma 3, we have the following result.

\begin{proposition}
Suppose that $f$ is a fair rank-minimizing mechanism. For $f$ and $a$ with
any $r_{a}\in \mathcal{P}$, truth-telling is not a dominated strategy.
\end{proposition}

This result implies that any fair rank-minimizing mechanism satisfies a weak
strategic characteristic; that is, for each untrue preference $p_{a}$,
either $\left( f\left( r_{a},p_{-a}\right) \right) _{ao}=\left( f\left(
p_{a},p_{-a}\right) \right) _{ao}$ for all $o\in \mathcal{O}$\ and all $%
p_{-a}\in \mathcal{P}^{\left\vert \mathcal{A}\right\vert -1}$ or there must
exist $p_{-a}$ such that $a$ is worse off by revealing it instead of $r_{a}$%
. Thus, if a fair rank-minimizing mechanism is employed, any (essential)
strategic manipulation by an agent carries some level of risk. This
characteristic is quite weak, but in the subsequent section, we show that
this is not satisfied when the agents can refuse their assignment and
instead obtain the null object type.

Note that Proposition 2 holds for any fair rank-minimizing mechanisms but
cannot be generalized to rank-minimizing mechanisms satisfying weak ETE. We
show this by introducing the following example.

\subsubsection*{Example 1}

Let $\mathcal{A=}\left\{ a_{1},a_{2},a_{3}\right\} $, $\mathcal{O=}\left\{
o_{1},o_{2},o_{3},\varnothing \right\} $ and $q_{o_{1}}=1$. Let $\alpha
=\left( o_{1},o_{2},\varnothing ,o_{3}\right) $ and $\alpha ^{\prime
}=\left( o_{1},o_{2},o_{3},\varnothing \right) $. Let a rank-minimizing
mechanism $f^{\ast }$ be such that, if $\left(
p_{a_{i}},p_{a_{j}},p_{a_{k}}\right) =\left( \alpha ^{\prime },\alpha
,\alpha \right) $ where $i,j,k=1,2,3$ and they are distinct, then 
\begin{eqnarray}
\left( f^{\ast }\left( p_{a_{1}},p_{a_{2}},p_{a_{3}}\right) \right)
_{a_{i}o_{1}} &=&0,  \label{g} \\
\left( f^{\ast }\left( p_{a_{1}},p_{a_{2}},p_{a_{3}}\right) \right) _{ao}
&=&\sum\limits_{y^{\ast }\in \mathcal{Y}^{\ast }\left( p\right) }\frac{%
y^{\ast }}{\left\vert \mathcal{Y}^{\ast }\left( p\right) \right\vert } 
\notag
\end{eqnarray}%
for all $a\in \mathcal{A}$ and $o\in \mathcal{O}$ otherwise; that is, if $%
\left( p_{a_{1}},p_{a_{2}},p_{a_{3}}\right) $ is neither $\left( \alpha
^{\prime },\alpha ,\alpha \right) $ nor $\left( \alpha ,\alpha ^{\prime
},\alpha \right) $ nor $\left( \alpha ,\alpha ,\alpha ^{\prime }\right) $.
In words, if the first-best type and the second-best type of the three
agents are respectively $o_{1}$ and $o_{2}$ and the third-best type of the
two of them is $\varnothing $, then the agent revealing $o_{3}$ as the
third-best one obtains $o_{1}$ with probability $0$.

We additionally assume that if $\left( p_{a_{i}},p_{a_{j}},p_{a_{k}}\right)
=\left( \alpha ^{\prime },\alpha ,\alpha \right) $, then 
\begin{equation*}
\left( f^{\ast }\left( p_{a_{1}},p_{a_{2}},p_{a_{3}}\right) \right) _{a_{j}%
\bar{o}}=\left( f^{\ast }\left( p_{a_{1}},p_{a_{2}},p_{a_{3}}\right) \right)
_{a_{k}\bar{o}}
\end{equation*}%
for all $\bar{o}\in \mathcal{O}$. Then, $f^{\ast }$ satisfies weak ETE.
However, whether $f^{\ast }$ satisfies ETE or not is dependent on $q_{o_{2}}$%
.

We assume that $r_{a_{1}}=\alpha ^{\prime }$ and this rank-minimizing
mechanism $f^{\ast }$ is employed. First, suppose $q_{o_{2}}=1$. In this
case, $f^{\ast }$ satisfies ETE and thus this is a fair rank-minimizing
mechanism. In this case, $\bar{k}\left( \alpha \right) =\bar{k}\left( \alpha
^{\prime }\right) =3$. Then, for example, $r_{a_{1}}=\alpha ^{\prime }$ is 
\textit{not} dominated by $\alpha $ and any other preference orders, because 
\begin{eqnarray*}
\left( f^{\ast }\left( \alpha ^{\prime },\alpha ^{\prime },\alpha ^{\prime
}\right) \right) _{a_{1}o_{1}} &=&\left( f^{\ast }\left( \alpha ^{\prime
},\alpha ^{\prime },\alpha ^{\prime }\right) \right) _{a_{1}o_{2}}=\left(
f^{\ast }\left( \alpha ^{\prime },\alpha ^{\prime },\alpha ^{\prime }\right)
\right) _{a_{1}o_{3}}=\frac{1}{3}, \\
\left( f^{\ast }\left( \alpha ,\alpha ^{\prime },\alpha ^{\prime }\right)
\right) _{a_{1}o_{1}} &=&\left( f^{\ast }\left( \alpha ,\alpha ^{\prime
},\alpha ^{\prime }\right) \right) _{a_{1}o_{2}}=\left( f^{\ast }\left(
\alpha ,\alpha ^{\prime },\alpha ^{\prime }\right) \right)
_{a_{1}\varnothing }=\frac{1}{3}.
\end{eqnarray*}%
That is, if $p_{a_{2}}=p_{a_{3}}=\alpha ^{\prime }$, then $a_{1}$ is worse
off by revealing $\alpha $ instead of $\alpha ^{\prime }=r_{a_{1}}\,$%
(truth-telling).

Second, suppose $q_{o_{2}}\geq 2$. In this case, $\bar{k}\left( \alpha
\right) =\bar{k}\left( \alpha ^{\prime }\right) =2$. Moreover, if $p_{a_{i}}$
is either $\alpha $ or $\alpha ^{\prime }$, then a rank-minimizing mechanism
satisfies 
\begin{equation*}
\left( f^{\ast }\left( p_{a_{1}},p_{a_{2}},p_{a_{3}}\right) \right)
_{a_{i}o_{1}}+\left( f^{\ast }\left( p_{a_{1}},p_{a_{2}},p_{a_{3}}\right)
\right) _{a_{i}o_{2}}=1.
\end{equation*}%
That is, if a rank-minimizing mechanism is employed and $a_{1}$ chooses
either $\alpha $ or $\alpha ^{\prime }$, then $a_{1}$ must have either $%
o_{1} $ or $o_{2}$. Thus, the only strategic interest of $a_{1}$ is
increasing the probability to obtain $o_{1}$. If $f^{\ast }$ is employed,
then $\alpha ^{\prime }=r_{a_{1}}$ is dominated by $\alpha $ for $a_{1}$ due
to (\ref{g}). Therefore, we conclude that even if a rank-minimizing
mechanism that satisfies weak ETE is employed, truth-telling may be
dominated by some other strategy. Note that in the same case, $f^{\ast }$
does not satisfy ETE. First, since $q_{o_{1}}+q_{o_{2}}\geq \left\vert 
\mathcal{A}\right\vert $, any agent is assigned to neither\ $o_{3}$ nor $%
\varnothing ,$ as long as the assignment is not wasteful. Thus, $\alpha $
and $\alpha ^{\prime }$ are essentially the same in this case. However, $%
\left( f^{\ast }\left( \alpha ^{\prime },\alpha ,\alpha \right) \right)
_{a_{1}o_{1}}=0,$ but $\left( f^{\ast }\left( \alpha ^{\prime },\alpha
,\alpha \right) \right) _{a_{1}o_{2}}>0$ or $\left( f^{\ast }\left( \alpha
^{\prime },\alpha ,\alpha \right) \right) _{a_{1}o_{3}}>0$ must be
satisfied. Hence $f^{\ast }$ does not satisfy ETE.

\section{Refusal option}

In this section, we consider the following \textit{two-stage game}, which is
a modification of that in Section 3. In the first stage, all agents
simultaneously reveal a preference order in $\mathcal{P}$ and an assignment
is determined via a mechanism based on $p$. In the second stage, all agents
decide whether to accept the assignment determined in the first stage. If an
agent refuses their assignment, then the agent instead obtains the null
object type $\varnothing $ (the outside option). That is, in this game, the
null object type has two characteristics: first, there are sufficiently many
copies of this type to allocate to all individuals, and second, agents can
always obtain it at no cost as long as they abandon their assignment.

This two-stage game is also considered by Featherstone (2020) but is not
explicitly addressed by almost other previous studies related to ours.%
\footnote{%
As exceptions, Feigenbaum et al. (2020) and Afacan (2022) consider an
assignment problem where, after the initial allocation is decided, if
someone refuses the assigned object, it is reallocated. Moreover, Do\u{g}an
and Yenmez (2019) also explicitly consider this stage.} However, this stage
is implicitly considered by many of them; that is, they consider an ex-post
individually rationality (or more strictly an ex-post efficiency) as a
requirement and if an assignment is ex-post individually rational, no agent
refuses their assignment. See, for example, Bogomolnaia and Moulin (2001) on
the property.

Moreover, note that, in this study, all agents are assumed to know the
capacities of all object types $\left( q_{o}\right) _{o\in \mathcal{O}}$,
when they reveal their preference. On the other hand, Featherstone (2020)
considers the situation where any agents do not know the capacity of any
object types. Due to this difference, some of our results differ from his.

In the third stage, the obvious optimal (dominant) strategy of an agent $a$
is to accept their assignment denoted by $o$ if $o$ acceptable; that is, if $%
R\left( o,r_{a}\right) <R\left( \varnothing ,r_{a}\right) $ and refuse it if
otherwise ($R\left( o,r_{a}\right) \geq R\left( \varnothing ,r_{a}\right) $%
). Based on this optimal strategy in the third stage, an assignment obtained
after the second stage denoted by $x$, and the profile of (true) preferences
denoted by $r$, we let $g\left( x,r\right) $ be such that for all $a\in 
\mathcal{A}$,%
\begin{eqnarray*}
g\left( x,r\right) _{ao} &=&x_{ao}\text{ for all }R\left( o,r_{a}\right)
<R\left( \varnothing ,r_{a}\right) \\
g\left( x,r\right) _{ao} &=&0\text{ for all }R\left( o,r_{a}\right) >R\left(
\varnothing ,r_{a}\right) \\
g\left( x,r\right) _{a\varnothing } &=&\sum\limits_{o:R\left( o,r_{a}\right)
\geq R\left( \varnothing ,r_{a}\right) }x_{ao}.
\end{eqnarray*}%
That is, since an agent refuses any of the unacceptable object types when it
is assigned, the probability that the null object type is assigned will
increase by the probability that any of the unacceptable object types is
assigned.

We reconsider strategic problems of a fair rank-minimizing mechanism in this
three-stage game. For a mechanism $f$ and an agent $a$, (revealing) $%
p_{a}^{\prime }$ is said to be \textbf{weakly} \textbf{dominated} \textbf{by 
}(revealing) $p_{a}$ \textbf{with the} \textbf{refusal option} if $a$ weakly
prefers $g\left( f\left( p_{a},p_{-a}\right) ,r\right) $ to $g\left( f\left(
p_{a}^{\prime },p_{-a}\right) ,r\right) $ for all $p_{-a}\in \mathcal{P}%
^{\left\vert \mathcal{A}\right\vert -1}$. Moreover, for $f$ and $a$,
(revealing) $p_{a}^{\prime }$ is said to be (strictly) \textbf{dominated} 
\textbf{by }(revealing) $p_{a}$ \textbf{with the} \textbf{refusal option }if 
$p_{a}^{\prime }$ is weakly dominated by $p_{a}$ with the refusal option%
\textbf{\ }and there exists some $p_{-a}^{\prime }\in \mathcal{P}%
^{\left\vert \mathcal{A}\right\vert -1}$ such that $a$ strictly prefers $%
g\left( f\left( p_{a},p_{-a}^{\prime }\right) ,r\right) $ to $g\left(
f\left( p_{a}^{\prime },p_{-a}^{\prime }\right) ,r\right) $. For a mechanism 
$f$ and an agent $a$, (revealing) $p_{a}\in \mathcal{P}$ is said to be a (%
\textbf{weakly}) \textbf{dominated strategy} \textbf{with the} \textbf{%
refusal option} if there is some $p_{a}^{\prime }\in \mathcal{P}$\ that
(weakly) dominates $p_{a}$ with the refusal option.

Moreover, we say that $p\in \mathcal{P}^{\left\vert \mathcal{A}\right\vert }$
is a \textbf{Nash equilibrium with the} \textbf{refusal option} if there is
no $a$ who strictly prefers $g\left( f\left( p_{a}^{\prime },p_{-a}\right)
,r\right) $ to $g\left( f\left( p_{a},p_{-a}\right) ,r\right) $ for some $%
p_{a}^{\prime }\in \mathcal{P}$.

We show that, under the URM, truth-telling is dominated by another strategy
when the agents have the refusal option. We define the following specific
strategies. Let $\mathcal{D}_{a}\subseteq \mathcal{P}$ be a set of
preferences such that $d_{a}\in \mathcal{D}_{a}$ if (i) $R\left(
o,r_{a}\right) =R\left( o,d_{a}\right) $ for all $o\in \mathcal{O}$ such
that $R\left( o,r_{a}\right) <R\left( \varnothing ,r_{a}\right) $ and (ii) $%
R\left( \varnothing ,d_{a}\right) =\left\vert \mathcal{O}\right\vert $.
First, the ranks of the truly acceptable object types are unchanged. Second,
the null object type $\varnothing $ is the least preferable one under $%
d_{a}\in \mathcal{D}_{a}$. We call the strategy of agent $a$ of revealing a
preference in $\mathcal{D}_{a}$ \textbf{an outside option demotion strategy }%
(hereafter\textbf{\ ODS})\textbf{\ }of agent $a$. Under an ODS, every object
type that is ranked below $\varnothing $ in the agent's true preference is
instead ranked above $\varnothing $ in $d_{a}$. Note that an ODS (revealing $%
d_{a}\in \mathcal{D}_{a}$) increases the number of acceptable object types
revealed by $a$; that is, $R\left( \varnothing ,r_{a}\right) \leq R\left(
\varnothing ,d_{a}\right) =\left\vert \mathcal{O}\right\vert $ is satisfied.

If $R\left( \varnothing ,r_{a}\right) =\left\vert \mathcal{O}\right\vert ,$
then truth-telling is the unique ODS for agent $a$. Therefore, an ODS
constitutes a strategic manipulation if and only if $R\left( \varnothing
,r_{a}\right) <\left\vert \mathcal{O}\right\vert ;$ that is, the null object
is not the worst object type for $a$.

Featherstone (2020) defines the \textbf{full extension} of $r_{a}\ $denoted
by $e_{a}$ in this study such that (i) $R\left( o,r_{a}\right) =R\left(
o,e_{a}\right) $ for all $o\in \mathcal{O}$ such that $R\left(
o,r_{a}\right) <R\left( \varnothing ,r_{a}\right) $, and (ii) $R\left(
\varnothing ,e_{a}\right) =\left\vert \mathcal{O}\right\vert $, and $R\left(
o,r_{a}\right) -1=R\left( o,e_{a}\right) $ for all $o\in \mathcal{O}$ such
that $R\left( o,r_{a}\right) >R\left( \varnothing ,r_{a}\right) $; that is, $%
e_{a}\in \mathcal{D}_{a}$ is such that all non-null object types are placed
in their true preference orders. Revealing the full extension of $r_{a}$ is
one of the ODSs of $a$.

First, we consider the following example to briefly consider the main result.

\subsubsection*{Example 2}

Let $\mathcal{A=}\left\{ a_{1},a_{2},a_{3}\right\} $ and $\mathcal{O=}%
\left\{ o_{1},o_{2},\varnothing \right\} $. Let $\beta =\left(
o_{1},o_{2},\varnothing \right) $, $\beta ^{\prime }=\left(
o_{2},o_{1},\varnothing \right) $ and $\beta ^{\prime \prime
}=(o_{1},\varnothing ,o_{2})$. Suppose $q_{o_{1}}=q_{o_{2}}=1$ and $%
r_{a_{1}}=\beta ^{\prime \prime }$. Then, the unique ODS is revealing $%
d_{a_{1}}=\beta $. In this example, we assume $\left(
p_{a_{2}},p_{a_{3}}\right) =\left( r_{a_{2}},r_{a_{3}}\right) ;$ that is, $%
a_{2}$ and $a_{3}$ always adopt truth-telling.

First, we consider the case where $p_{a_{2}}=\beta $ and $p_{a_{3}}=\beta
^{\prime }$. Then, 
\begin{equation*}
f^{U}\left( r_{a_{1}},\beta ,\beta ^{\prime }\right) =y=\left( 
\begin{array}{ccc}
y_{a_{1}o_{1}} & y_{a_{1}o_{2}} & y_{a_{1}\varnothing } \\ 
y_{a_{2}o_{1}} & y_{a_{2}o_{2}} & y_{a_{2}\varnothing } \\ 
y_{a_{3}o_{1}} & y_{a_{3}o_{2}} & y_{a_{3}\varnothing }%
\end{array}%
\right) =\left( 
\begin{array}{ccc}
0 & 0 & 1 \\ 
1 & 0 & 0 \\ 
0 & 1 & 0%
\end{array}%
\right) .
\end{equation*}%
Whenever $\left\vert \mathcal{A}\right\vert =\left\vert \mathcal{O}%
\right\vert =3$, we write the matrix representing an assignment in this
manner. On the other hand, 
\begin{equation*}
f^{U}\left( d_{a_{1}},\beta ,\beta ^{\prime }\right) =\left( 
\begin{array}{ccc}
1/2 & 0 & 1/2 \\ 
1/2 & 0 & 1/2 \\ 
0 & 1 & 0%
\end{array}%
\right)
\end{equation*}%
Thus, revealing $d_{a_{1}}$ instead of $r_{a_{1}}$ may increase the
probability that $a_{1}$ is assigned to their first-best object $o_{1}$.
First, under $\left( r_{a_{1}},\beta ,\beta ^{\prime }\right) ,$ $%
\varnothing $ is ranked lowest by $a_{2}$ and $a_{3}$ but $a_{1}$ ranks $%
\varnothing $ as their second-best. Thus, in any rank-minimizing mechanism, $%
a_{1}$ is assigned $\varnothing $ with probability 1. On the other hand,
since $a_{1}$ places $\varnothing $ at the bottom of their ranking. As a
result. under $\left( d_{a_{1}},p_{a_{2}},p_{a_{3}}\right) $, $a_{1}$ is
assigned $o_{1}$---the object that is truly their best---with probability $%
1/2$, and $\varnothing $ with probability $1/2$. Thus, in this case, $a_{1}$
is better off by choosing $d_{a_{1}}$ instead of $r_{a_{1}}$.

Second, suppose $p_{a_{2}}=p_{a_{3}}=\beta ^{\prime \prime }.$ Then, 
\begin{gather*}
f^{U}\left( r_{a_{1}},\beta ^{\prime \prime },\beta ^{\prime \prime }\right)
=\left( 
\begin{array}{ccc}
1/3 & 0 & 2/3 \\ 
1/3 & 0 & 2/3 \\ 
1/3 & 0 & 2/3%
\end{array}%
\right) , \\
f^{U}\left( d_{a_{1}},\beta ^{\prime \prime },\beta ^{\prime \prime }\right)
=\left( 
\begin{array}{ccc}
1/3 & 2/3 & 0 \\ 
1/3 & 0 & 2/3 \\ 
1/3 & 0 & 2/3%
\end{array}%
\right) .
\end{gather*}%
Thus, in this case, $a_{1}$ seems to be worse off by revealing $d_{a_{1}}$
instead of $r_{a_{1}}$, because $a_{1}$ prefers $\varnothing $ to $o_{2}$.
However, $a_{1}$ can also refuse $o_{2}$ and obtain $\varnothing $.
Therefore, if $r_{a_{2}}=r_{a_{3}}=\beta ^{\prime \prime }$,%
\begin{equation*}
g\left( f^{U}\left( d_{a_{1}},\beta ^{\prime \prime },\beta ^{\prime \prime
}\right) ,r\right) =g\left( f^{U}\left( r_{a_{1}},\beta ^{\prime \prime
},\beta ^{\prime \prime }\right) ,r\right) =f^{U}\left( r_{a_{1}},\beta
^{\prime \prime },\beta ^{\prime \prime }\right) \text{;}
\end{equation*}%
that is, $a_{1}$ is not worse off by choosing $d_{a_{1}}$ instead of $%
r_{a_{1}}$.

Thus, in this example, $a_{1}$ is never worse off and may even be better off
by selecting $d_{a_{1}}$ (the ODS) instead of $r_{a_{1}}$. Next, we show
that the ODS by $a_{1}$ may make the assignment wasteful.

Third, suppose $p_{a_{2}}=p_{a_{3}}=\beta $. Then, 
\begin{gather}
f^{U}\left( r_{a_{1}},\beta ,\beta \right) =\left( 
\begin{array}{ccc}
0 & 0 & 1 \\ 
1/2 & 1/2 & 0 \\ 
1/2 & 1/2 & 0%
\end{array}%
\right) ,  \notag \\
f^{U}\left( d_{a_{1}},\beta ,\beta \right) =\left( 
\begin{array}{ccc}
1/3 & 1/3 & 1/3 \\ 
1/3 & 1/3 & 1/3 \\ 
1/3 & 1/3 & 1/3%
\end{array}%
\right) .  \label{y1}
\end{gather}%
In the latter case, $a_{1}$ refuses the assignment; that is, 
\begin{equation}
g\left( f^{U}\left( d_{a_{1}},\beta ,\beta \right) ,r\right) =\left( 
\begin{array}{ccc}
1/3 & 0 & 2/3 \\ 
1/3 & 1/3 & 1/3 \\ 
1/3 & 1/3 & 1/3%
\end{array}%
\right) .  \label{y2}
\end{equation}%
Then, $a_{1}$ is better off by adopting the ODS instead of truth-telling.
However, $g\left( f^{U}\left( d_{a_{1}},\beta ,\beta \right) ,r\right) $ is
wasteful, because 
\begin{eqnarray*}
\sum\limits_{a\in \left\{ a_{1},a_{2},a_{3}\right\} }g\left( f^{U}\left(
d_{a_{1}},\beta ,\beta \right) ,r\right) _{ao_{2}} &=&\frac{2}{3}<1, \\
\text{and }g\left( f^{U}\left( d_{a_{1}},\beta ,\beta \right) ,r\right)
_{a^{\prime }\varnothing } &=&\frac{1}{3}>0,
\end{eqnarray*}%
for $a^{\prime }=a_{2},a_{3}$. That is, for 
\begin{equation*}
\tilde{x}=\left( 
\begin{array}{ccc}
1/3 & 0 & 2/3 \\ 
1/3 & 1/2 & 1/6 \\ 
1/3 & 1/2 & 1/6%
\end{array}%
\right) ,
\end{equation*}%
$a_{2}$ and $a_{3}$ strictly prefer and $a_{1}$ weakly prefers $\tilde{x}$
to $g\left( f^{U}\left( d_{a_{1}},\beta ,\beta \right) ,r\right) $.

If $a_{1}$ is assigned to $o_{2}$, then either $a_{2}$ or $a_{3}$ will be
assigned to $\varnothing $. However, in this case, $a_{1}$ will refuse $%
o_{2} $ and thus $o_{2}$ will remain unallocated. In this case, rather than
leaving $o_{2}$ unallocated, it is more efficient to assign $o_{2}$ to the
agent ($a_{2}$ or $a_{3}$) who would otherwise receive $\varnothing $ and
instead assign $a_{1}$ to $\varnothing $ from the beginning. This adjustment
ensures that $o_{2}$ is allocated efficiently. Hence revealing $d_{a_{1}}$
and refusing an unacceptable assignment may benefit $a_{1}$ but lead to a
wasteful allocation, as some agents may prefer the object type that $a_{1}$
refuses. \newline

Now, we formally introduce our main results.

\begin{theorem}
For $f^{U}$ and any agent $a$, truth-telling is weakly dominated strategy;
that is, it is weakly dominated by any ODS with the refusal option.
\end{theorem}

By this result, an agent is never hurt by using an ODS instead of
truth-telling. Next, we show that with some conditions, there is some ODS
such that an agent may be better off by using it instead of truth-telling.

\begin{theorem}
Suppose that there is $a\in \mathcal{A}$ whose true preference $r_{a}$
satisfies 
\begin{equation}
R\left( o,r_{a}\right) <R\left( \varnothing ,r_{a}\right) <R\left( o^{\prime
},r_{a}\right) \text{ and }\sum\limits_{\bar{o}:R\left( \bar{o},r_{a}\right)
<R\left( \varnothing ,r_{a}\right) }q_{\bar{o}}+q_{o^{\prime }}<\left\vert 
\mathcal{A}\right\vert \text{,}  \label{f}
\end{equation}%
for some $o,o^{\prime }\in \mathcal{O}$. Let$\ d_{a}\in \mathcal{D}_{a}$
such that $R\left( \varnothing ,r_{a}\right) =R\left( o^{\prime
},d_{a}\right) $. Then, for $f^{U}$ and $a$, truth-telling is a dominated
strategy; that is, it is dominated by an ODS of revealing$\ d_{a}\in 
\mathcal{D}_{a}$ with the refusal option.
\end{theorem}

This result implies that if there is $a\in \mathcal{A}$ satisfying (\ref{f})
for some $o,o^{\prime }\in \mathcal{O}$, then agent $a$ has a strong
incentive to adopt an ODS. This is satisfied only when some agent has at
least one unacceptable object. The condition given in (\ref{f}) is that
there is a truly unacceptable type $o^{\prime }$ such that the number of
copies of $o^{\prime }$, plus the sum of the copies of the truly acceptable
types, is sufficiently small. We discuss real-world scenarios where (\ref{f}%
) is satisfied for some $a,o,o^{\prime }$ in the subsequent section.

Theorems 1 and 2 do not imply that an ODS is a dominant strategy. However,
if there is $a\in \mathcal{A}$ whose true preference satisfies (\ref{f}) for
some $o,o^{\prime }\in \mathcal{O}$, since truth-telling is a dominated
strategy, $a$ has a strong incentive to manipulate its preference ranking.

As stated earlier, multiple ODSs may exist. Even if an ODS dominates
truth-telling, another ODS that does not dominate truth-telling may also
exist. In such a case, the latter ODS is dominated by the former. The
strategy of revealing the full extension is not an exception to this. We
show this fact by introducing the following example.

\subsubsection*{Example 3}

Let $\mathcal{A=}\left\{ a_{1},a_{2},a_{3}\right\} $ and $\mathcal{O=}%
\left\{ o_{1},o_{2},o_{3},\varnothing \right\} $. Let $\gamma
=(o_{1},o_{2},o_{3},\varnothing ),$ $\gamma ^{\prime
}=(o_{1},o_{3},o_{2},\varnothing )$ and $\gamma ^{\prime \prime
}=(o_{3},o_{1},o_{2},\varnothing ).$ Suppose $q_{o_{1}}=q_{o_{3}}=1,$ $%
q_{o_{2}}=2$ and $r_{a_{1}}=(o_{1},\varnothing ,o_{2},o_{3})$. Then, there
are two ODSs. One is $e_{a_{1}}=\gamma $, which is the full extension of $%
a_{1}$, and the other is $d_{a_{1}}=\gamma ^{\prime }$. In this example, we
assume $\left( p_{a_{2}},p_{a_{3}}\right) =\left( r_{a_{2}},r_{a_{3}}\right)
;$ that is, $a_{2}$ and $a_{3}$ always adopt truth-telling.

First, suppose $p_{a_{2}}=\gamma ^{\prime }$ and $p_{a_{3}}=\gamma ^{\prime
\prime }.$ Then, 
\begin{equation*}
f^{U}\left( r_{a_{1}},\gamma ^{\prime },\gamma ^{\prime \prime }\right)
=y=\left( 
\begin{array}{cccc}
y_{a_{1}o_{1}} & y_{a_{1}o_{2}} & y_{a_{1}o_{3}} & y_{a_{1}\varnothing } \\ 
y_{a_{2}o_{1}} & y_{a_{2}o_{2}} & y_{a_{2}o_{3}} & y_{a_{2}\varnothing } \\ 
y_{a_{3}o_{1}} & y_{a_{3}o_{2}} & y_{a_{3}o_{3}} & y_{a_{3}\varnothing }%
\end{array}%
\right) =\left( 
\begin{array}{cccc}
0 & 0 & 0 & 1 \\ 
1 & 0 & 0 & 0 \\ 
0 & 0 & 1 & 0%
\end{array}%
\right) ,
\end{equation*}%
In this example, we write the assignments in this manner. Moreover, 
\begin{equation*}
g\left( f^{U}\left( d_{a_{1}},\gamma ^{\prime },\gamma ^{\prime \prime
}\right) ,r\right) =\left( 
\begin{array}{cccc}
1/2 & 0 & 0 & 1/2 \\ 
1/2 & 1/2 & 0 & 0 \\ 
0 & 0 & 1 & 0%
\end{array}%
\right) \text{.}
\end{equation*}%
Therefore, $a_{1}$ (strictly) prefers $g\left( f^{U}\left( d_{a_{1}},\gamma
^{\prime },\gamma ^{\prime \prime }\right) ,r\right) $ to $g\left(
f^{U}\left( r_{a_{1}},\gamma ^{\prime },\gamma ^{\prime \prime }\right)
,r\right) $ and $r_{a_{1}}$ is dominated by $d_{a_{1}}$.

On the other hand, there is no pair $\left( p_{a_{2}},p_{a_{3}}\right) $ of
preference orders of $a_{2}$ and $a_{3}$ such that $a_{1}$ (strictly)
prefers $g\left( f^{U}\left( e_{a_{1}},p_{a_{2}},p_{a_{3}}\right) ,r\right) $
to $g\left( f^{U}\left( r_{a_{1}},p_{a_{2}},p_{a_{3}}\right) ,r\right) $.
This is because, by $q_{o_{1}}+q_{o_{2}}=3=\left\vert \mathcal{A}\right\vert 
$, for any $\left( p_{a_{2}},p_{a_{3}}\right) \in \mathcal{P}^{2}$ and any $%
\bar{o}\in \mathcal{O}$, 
\begin{equation*}
\left( g\left( f^{U}\left( e_{a_{1}},p_{a_{2}},p_{a_{3}}\right) ,r\right)
\right) _{a_{1}\bar{o}}=\left( g\left( f^{U}\left(
r_{a_{1}},p_{a_{2}},p_{a_{3}}\right) ,r\right) \right) _{a_{1}\bar{o}}\text{.%
}
\end{equation*}%
Thus, $e_{a_{1}}$ is also dominated by $d_{a_{1}}$.

We intuitively explain why $d_{a_{1}}$ dominates $e_{a_{1}}$. In this
example, $q_{o_{1}}+q_{o_{3}}=2<3=q_{o_{1}}+q_{o_{2}}$. That is, the copies
of $o_{1}$ and $o_{2}$ are sufficient to allocate all agents, but those of $%
o_{1}$ and $o_{3}$ are insufficient. Thus, revealing $o_{2}$ as a
second-best type does not increase the probability that $a_{1}$ is assigned
to $o_{1}$. On the other hand, revealing $o_{3}$ as a second-best type
increases it, when $o_{3}$ is the first-best type of another agent. This is
because, to minimize the total rank, $o_{3}$ must be assigned to an agent
who ranks it as their first-best type. However, since $a_{1}$ ranks $o_{2}$
and $\varnothing $ lower, minimizing the total rank requires increasing the
probability that $o_{1}$ is allocated to $a_{1}$. Thus, revealing the full
extension may be dominated by another ODS when an unacceptable type (in this
case $o_{3}$) is less preferable than the other (in this case $o_{2}$) but
the number of copies of the former is less than that of the latter.\newline

Note that this fact does not contradict a result of Featherstone (2020,
Proposition 13). In their model, they assume that the agents have no
information on the capacities of object types. Revealing the full extension
can be dominated by another ODS only when the capacities of the unacceptable
types are different and agents know them. Hence the strategic domination in
Example 3 does not occur in the low-information situation discussed by
Featherstone (2020).

Next, we show that if an agent adopts an ODS, the resulting assignment may
be wasteful and thus inefficient. Moreover, such an outcome can arise in a
Nash equilibrium.

\begin{proposition}
There exist $r\in \mathcal{P}^{\left\vert \mathcal{A}\right\vert }$ and $%
d\in \mathcal{P}^{\left\vert \mathcal{A}\right\vert }$ such that $d,$ the
profile in which all agents adopt ODSs, is a Nash equilibrium with the
refusal option under $f^{U}$. Moreover, if there are $o,o^{\prime }\in 
\mathcal{O}$ such that $q_{o}+q_{o^{\prime }}<\left\vert \mathcal{A}%
\right\vert $, then such profiles $r\in \mathcal{P}^{\left\vert \mathcal{A}%
\right\vert }$ and $d\in \mathcal{P}^{\left\vert \mathcal{A}\right\vert }$
can also yield a wasteful assignment; that is, $g\left( f^{U}\left( d\right)
,r\right) $ is wasteful.
\end{proposition}

We briefly show this by revisiting Example 2. Let 
\begin{equation*}
r=\left( r_{a_{1}},r_{a_{2}},r_{a_{3}}\right) =\left( \beta ^{\prime \prime
},\beta ,\beta \right) \text{ and }d=\left(
d_{a_{1}},d_{a_{2}},d_{a_{3}}\right) =\left( \beta ,\beta ,\beta \right) .
\end{equation*}%
Then, $g\left( f^{U}\left( d\right) ,r\right) $ is given by (\ref{y2}),
which is a wasteful assignment. Moreover, $d=\left( \beta ,\beta ,\beta
\right) $ is a Nash equilibrium; that is, for each agent $a$, $g\left(
f^{U}\left( d\right) ,r\right) $ is not first-order stochastically dominated
by $g\left( f^{U}\left( p_{a},d_{-a}\right) ,r\right) $ for any $p_{a}\in 
\mathcal{P}$.\footnote{%
Since we consider the first-order stochastic dominance relation, our notion
of Nash equilibrium is weaker than the standard one. Nevertheless, suppose
that both $a_{2}$ and $a_{3}$ prefer the assignment such that each object $%
o\in \left\{ o_{1},o_{2},\varnothing \right\} $ assigns with the probability 
$1/3$ to the assignment such that $o_{2}$ assigns with probability $1$.\
This assumption ensures that neither $a_{2}$ nor $a_{3}$ has an incentive to
change $\beta $ and $\beta $ to $(o_{2},o_{1},\varnothing )$ or $%
(o_{2},\varnothing ,o_{1})$. Under this condition, $d=\left( \beta ,\beta
,\beta \right) $ is a Nash equilibrium even in the standard sense.}

By Proposition 3, if agents adopt an ODS, then the result may be wasteful.
Moreover, such a strategy profile can be a Nash equilibrium. Therefore,
under the URM, a profile of ODS are stably adopted in place of truth-telling
and that such strategies may result in inefficient outcomes. This suggests a
significant drawback of the URM.

Therefore, we next consider an alternative fair rank-minimizing mechanism
which, unlike $f^{U}$, guarantees that for any agent, truth-telling is 
\textit{not} dominated by any other strategy when it is used as the
mechanism in the second stage.

We let $f^{M}$ be the \textbf{modified uniform rank-minimizing mechanism }%
(hereafter modified URM) such that if $p$ satisfies

\begin{description}
\item[(i)] there is one agent represented by $a$ such that $R\left(
o,p_{a}\right) =1$ and $R\left( \varnothing ,p_{a}\right) \geq 3,$

\item[(ii)] there are $q_{o}$ agents represented by $a^{\prime }$ revealing $%
p_{a^{\prime }}$, such that $R\left( \varnothing ,p_{a^{\prime }}\right)
=l\geq 2,$ $p_{a}^{l^{\prime }}=p_{a^{\prime }}^{l}$ for all $l^{\prime
}=1,\cdots ,l-1,$ $R\left( \varnothing ,p_{a}\right) >l,$ and $\bar{k}%
(p_{a^{\prime }})=l$; that is,%
\begin{equation*}
\sum\limits_{\bar{o}:R\left( \bar{o},p_{a^{\prime }}\right) <R\left(
\varnothing ,p_{a^{\prime }}\right) }q_{\bar{o}}<\left\vert \mathcal{A}%
\right\vert ,
\end{equation*}

\item[(iii)] if $\left\vert \mathcal{A}\right\vert >q_{o}+1$, then the
first-best type of the other agents ($\left\vert \mathcal{A}\right\vert
-\left( q_{o}+1\right) $ agents) represented by $a^{\prime \prime }$ is $%
\varnothing $, then
\end{description}

\begin{equation*}
\left( f^{M}\left( p\right) \right) _{ao}=0,\text{ }\left( f^{M}\left(
p\right) \right) _{a^{\prime }o}=1,\text{ }\left( f^{M}\left( p\right)
\right) _{ap_{a}^{2}}=1\text{ and }\left( f^{M}\left( p\right) \right)
_{a^{\prime \prime }\varnothing }=1,
\end{equation*}%
and otherwise; that is, if $p$ does not satisfy all of them, then 
\begin{equation*}
f^{M}\left( p\right) =\sum\limits_{y^{\ast }\in \mathcal{Y}^{\ast }\left(
p\right) }\frac{y^{\ast }}{\left\vert \mathcal{Y}^{\ast }\left( p\right)
\right\vert }=f^{U}\left( p\right) .
\end{equation*}

Under this mechanism, the outcome is equivalent to that of the URM, except
when a specific preference profile satisfying conditions (i)--(iii) is
realized. The specific profile characterized by conditions (i)--(iii) is one
in which: (i) a particular agent denoted by $a$ ranks object $o$ first and
the outside option at position three or lower; (ii) there are $q_{o}$ other
agents whose preferences match $a$'s for the top $l-1$ positions, rank the
outside option exactly $l$th, and for whom the total capacity of objects
ranked above $l$ is insufficient for all agents; and (iii) if other\ agents
remain, all of them rank the outside option first.

We have the following result.

\begin{proposition}
The modified URM satisfies ETE.
\end{proposition}

Then, we explain why the strategic issues, mentioned in Theorems 1 and 2, do
not arise under this mechanism $f^{M}$. To ensure that truth-telling is not
dominated by any ODS, the mechanism must be such that, for every agent and
ODS, there exists a strategy profile of the other agents under which the
agent is not assigned to any of the more preferred object types. Suppose
that the true preference of $a$ satisfies $r_{a}^{1}=o$ and $%
r_{a}^{2}=\varnothing $. Given $p_{-a}$ satisfying conditions (ii) and
(iii), condition (i) is not satisfied if agent $a$ chooses truth-telling,
but it is satisfied if $a$ adopts an ODS. Therefore, when agent $a$ adopts
an ODS, the probability of being assigned their most preferred object $o$
becomes zero, whereas under truth-telling, this probability is positive.

To understand this point more, we revisit Example \ 2. In this example, $%
\left( d_{a_{1}},p_{a_{2}},p_{a_{3}}\right) =\left( \beta ,\beta ^{\prime
\prime },\beta ^{\prime \prime }\right) $ satisfies (i), (ii) and (iii), but 
$\left( r_{a_{1}},p_{a_{2}},p_{a_{3}}\right) =\left( \beta ^{\prime \prime
},\beta ^{\prime \prime },\beta ^{\prime \prime }\right) $ does not. Thus, 
\begin{eqnarray*}
f^{M}\left( r_{a_{1}},\beta ^{\prime \prime },\beta ^{\prime \prime }\right)
&=&f^{U}\left( r_{a_{1}},\beta ^{\prime \prime },\beta ^{\prime \prime
}\right) =\left( 
\begin{array}{ccc}
1/3 & 0 & 2/3 \\ 
1/3 & 0 & 2/3 \\ 
1/3 & 0 & 2/3%
\end{array}%
\right) , \\
f^{M}\left( d_{a_{1}},\beta ^{\prime \prime },\beta ^{\prime \prime }\right)
&=&\left( 
\begin{array}{ccc}
0 & 1 & 0 \\ 
1/2 & 0 & 1/2 \\ 
1/2 & 0 & 1/2%
\end{array}%
\right) , \\
g\left( f^{M}\left( d_{a_{1}},\beta ^{\prime \prime },\beta ^{\prime \prime
}\right) \right) &=&\left( 
\begin{array}{ccc}
0 & 0 & 1 \\ 
1/2 & 0 & 1/2 \\ 
1/2 & 0 & 1/2%
\end{array}%
\right) .
\end{eqnarray*}%
This implies that $a_{1}$ may be worse off by revealing $d_{a_{1}}$ instead
of $r_{a_{1}}$, even if $a_{1}$ refuses any unacceptable objects. Therefore, 
$r_{a_{1}}$ is not dominated by $d_{a_{1}}$, if $f^{M}$ is employed.

Formally, we have the following fact.

\begin{proposition}
For $f^{M}$ and $a$ with any $r_{a}\in \mathcal{P}$, truth-telling is not a
dominated strategy with the refusal option.
\end{proposition}

By adopting $f^{M}$, any ODS can fail because increasing the number of
acceptable objects may lead the agent to be unassigned to truly acceptable
objects, depending on the strategy profile of the other agents. Thus, in
some profiles of other agents' (revealed) preferences, truth-telling is
better than adopting any ODS.

However, the modified URM may introduce alternative strategic concerns. We
now turn to a different class of strategic manipulations.

Let $\Pi _{a}\subseteq \mathcal{P}$ be a set of preferences such
that $\pi _{a}\in \Pi _{a}$ if (i) $R\left( o,r_{a}\right)
=R\left( o,\pi _{a}\right) $ for all $o\in \mathcal{O}$ such that $R\left(
o,\pi _{a}\right) >R\left( \varnothing ,\pi _{a}\right) $, (ii) $R\left(
\varnothing ,\pi _{a}\right) <R\left( \varnothing ,r_{a}\right) $, and (iii) 
$R\left( o,r_{a}\right) =R\left( o,\pi _{a}\right) +1$ for all $o$ such that 
$R\left( \varnothing ,\pi _{a}\right) \leq R\left( o,r_{a}\right) <R\left(
\varnothing ,r_{a}\right) $, and (iv) $R\left( o,r_{a}\right) =R\left( o,\pi
_{a}\right) $ for all $o$ such that $R\left( \varnothing ,r_{a}\right)
<R\left( o,r_{a}\right) $. We call the strategy of agent $a$ of revealing a
preference in $\Pi _{a}$ \textbf{an outside option promotion
strategy }(hereafter\textbf{\ OPS})\textbf{\ }of agent $a$. In words,
contrary to ODSs, an agent raises the rank of the outside option in their
reported preference, making fewer object types appear acceptable.

First, we have the following result on this class of strategies on the URM
(not modified one).

\begin{proposition}
For $f^{M}$ and $a$ with any $r_{a}\in \mathcal{P}$, any OPS of $a$ is
weakly dominated by truth-telling of $a$. Moreover, this is satisfied even
without the refusal option.
\end{proposition}

On the other hand, if the modified URM is adopted instead of the URM, an
agent may have an incentive to adopt an OPS. We briefly show this fact by
using the following example.

\subsubsection*{Example 4}

Let $\mathcal{A=}\left\{ a_{1},a_{2},a_{3}\right\} $ and $\mathcal{O=}%
\left\{ o_{1},o_{2},\varnothing \right\} $. Let $\delta =\left(
o_{1},o_{2},\varnothing \right) $, $\delta ^{\prime }=\left(
o_{1},\varnothing ,o_{2}\right) $ and $\delta ^{\prime \prime }=\left(
\varnothing ,o_{1},o_{2}\right) $. We assume $q_{o_{1}}=q_{o_{2}}=1$.
Suppose $r_{a_{1}}=\delta $, $p_{a_{2}}=\delta $ and $p_{a_{3}}=\delta
^{\prime \prime }$. Then, $\delta ^{\prime }$ is an OPS of $a_{1}$. In this
case, since $\left( \delta (=r_{a_{1}}),\delta \right) $ does not satisfy
all of (i), (ii) and (iii),%
\begin{eqnarray*}
\left( f^{M}\left( \delta ,\delta ,\delta ^{\prime \prime }\right) \right)
_{a_{1}o_{1}} &=&\left( f^{M}\left( \delta ,\delta ,\delta ^{\prime \prime
}\right) \right) _{a_{1}o_{2}}=\frac{1}{2}, \\
\left( f^{U}\left( \delta ,\delta ,\delta ^{\prime \prime }\right) \right)
_{a_{1}o_{1}} &=&\left( f^{U}\left( \delta ,\delta ,\delta ^{\prime \prime
}\right) \right) _{a_{1}o_{2}}=\frac{1}{2}.
\end{eqnarray*}%
On the other hand, since $\left( \delta ^{\prime }(\neq r_{a_{1}}),\delta
\right) $ satisfies all of (i), (ii) and (iii),%
\begin{eqnarray*}
\left( f^{M}\left( \delta ^{\prime },\delta ,\delta ^{\prime \prime }\right)
\right) _{a_{1}o_{1}} &=&1,\text{ }\left( f^{M}\left( \delta ^{\prime
},\delta ,\delta ^{\prime \prime }\right) \right) _{a_{1}o_{2}}=0, \\
\left( f^{U}\left( \delta ^{\prime },\delta ,\delta ^{\prime \prime }\right)
\right) _{a_{1}o_{1}} &=&\left( f^{U}\left( \delta ^{\prime },\delta ,\delta
^{\prime \prime }\right) \right) _{a_{1}o_{2}}=\frac{1}{2}.
\end{eqnarray*}%
Thus, in this case, if $f^{M}$ is employed, then $a_{1}$ has an incentive to
choose $\delta ^{\prime }$ instead of $r_{a_{1}}=\delta $. In other words,
in this case, $a_{1}$ has an incentive to use an OPS. On the other hand, if $%
f^{U}$ is employed, then $a_{1}$ has no such an\ incentive.\newline

Formally, we have the following result.

\begin{proposition}
Let $f$ be a fair rank-minimizing rule such that, for any $r_{a}\in \mathcal{%
P}$, there is no $p_{a}\in \mathcal{P}$ that dominates $r_{a}$ with a
refusal option. Then, there is some $(r_{a},p_{-a})$ such that $a$ strictly
prefers $g\left( f\left( \pi _{a},p_{-a}\right) ,r\right) $ to $g\left(
f\left( r_{a},p_{-a}\right) ,r\right) $ where $\pi _{a}$ is an OPS of $a$.
Moreover, this is satisfied even without the refusal option; that is, for
that $(r_{a},p_{-a})$, $a$ strictly prefers $f\left( \pi _{a},p_{-a}\right) $
to $f\left( r_{a},p_{-a}\right) $.
\end{proposition}

Proposition 5 shows that it is possible to construct a fair rank-minimizing
mechanism under which truth-telling is not a dominated strategy. However,
doing so gives rise to a new type of strategic concern that does not exist
under the URM. Specifically, as shown in Proposition 6, under the URM, any
OPS is dominated by truth-telling, thereby eliminating the risk of such
manipulation. In contrast, Proposition 7 implies that under a fair
rank-minimizing mechanism in which truth-telling is not dominated by any
strategies, including the modified URM, an OPS is not dominated by
truth-telling; that is, for an agent, there exsits some profile of other
agents This suggests that manipulation through such strategies may persist
even in mechanisms designed to eliminate strategic dominance of
truth-telling.

Of course, this does not mean that truth-telling is dominated by an OPS.
However, this reveals a trade-off inherent in fair rank-minimizing
mechanisms: while modifying them can eliminate the strategic dominance of
truth-telling, such modifications may introduce incentives for other
manipulations.

\section{Concluding Remarks}

We show that if an assignment is determined by the URM, which is a fair one,
and agents can refuse the assignment and obtain the outside option instead,
then truth-telling is dominated by an ODS. The adoption of such strategies
by agents may result in inefficient assignments; that is, an agent who
adopts an ODS refuses their assigned type, although some other agents desire
that type.

We consider a public school choice system as an example. Our result implies
that a student may have an incentive to reveal truly unacceptable schools as
acceptable schools, and they may be assigned to one of them and then refuse
the admission even if some other students desire the school. In such a case,
the school might re-enroll students, but even in that case, it would cause
market disruption.

Our result implies that if all agents prefer many other types over the
outside option, then the strategic problem considered in this study is not
so severe. For example, in the case of matching systems for schools that are
part of compulsory education, the outside option does not need to be
considered. Therefore, there is no strategic problem of the kind discussed
in this study. On the other hand, the strategic problem may be severe when
there are agents who relatively highly value the outside option.

Moreover, if agents who refuse their assignment can be sufficiently
penalized, the problem would be resolved. However, considering that some
agents may refuse their assignment due to unavoidable circumstances (without
any strategic intention), the penalties should not be too severe.

Therefore, if resolving the problem in that way is difficult, it is
necessary to consider alternative mechanisms. First, as shown in this study,
if we modify the URM, then truth-telling is not dominated by any other
strategies, including the outside option demotion ones. However, in this
modification, it is necessary to ensure that increasing the number of
acceptable types can result in a disadvantage. Thus, this modification leads
to the opposite strategic problem, where agents have an incentive to
decrease the number of acceptable types.

Second, we consider efficient and fair mechanisms other than the fair
rank-minimizing mechanisms, such as the (general) probabilistic serial
mechanism introduced by Bogomolnaia and Moulin (2001) and Budish et al.
(2013), because it is fair and ordinally efficient, although this efficiency
property is weaker than the rank-minimizing.

Finally, our study does not address reallocation after assignment refusals.
However, Feigenbaum et al. (2020) and Afacan (2022) consider dynamic
markets, such as school choice markets with a second-round admission
process, where reallocation after refusals is possible. Incorporating such a
framework could enhance the practical applicability of our findings.
Specifically, analyzing how refused objects are redistributed and how
agents' strategic behavior changes in response could provide new insights
into the design of rank-minimizing mechanisms.

\section*{References}

\begin{description}
\item Afacan, M.O. 2022. Two-Stage Assignments with the Outside Alternative,
Available at SSRN: https://ssrn.com/abstract=4068725

\item Bando, K., Kokubo, N., Matsui, T. 2025. A Polynomial-time Algorithm
for Strategic Manipulations in Weight-minimizing Assignment Mechanisms,
Available at SSRN: https://ssrn.com/abstract=5053175

\item Bogomolnaia, A., Moulin, H. 2001. A New Solution to the Random
Assignment Problem. Journal of Economic Theory 100, 295-328.

\item Budish, E., Che, Y.-K., Kojima, F., Milgrom, P. 2013. Designing Random
Allocation Mechanisms: Theory and Applications. American Economic Review
103(2), 585--623.

\item Do\u{g}an, B., Yenmez, M.B. 2019. Unified versus Divided Enrollment in
School Choice: Improving Student Welfare in Chicago. Games and Economic
Behavior 118, 366-373

\item Featherstone, C. 2020. Rank efficiency: Modeling a common policymaker
objective. University of Pennsylvania. Unpublished paper.

\item Feigenbaum, I., Kanoria, Y., Lo, I., Sethuraman, J. 2020. Dynamic
Matching in School Choice: Efficient Seat Reassignment after Late
Cancellations, Management Science 66(11), 5341-5361.

\item Feizi, M. 2024. Notions of Rank Efficiency for the Random Assignment
Problem. Journal of Public Economic Theory 26(6), e70008

\item Han, X. 2024. On the efficiency and fairness of deferred acceptance
with single tie-breaking. Journal of Economic Theory 218, 105842

\item Kojima, F., Manea, M. 2010. Incentives in the probabilistic serial
mechanism. Journal of Economic Theory 145(1), 106-123.

\item  Krokhmal, A.P., Pardalos, P.M. 2009. Random assignment problems. European Journal of Operational Research, 194(1) 1-17    

\item Nesterov, A.S. 2017. Fairness and efficiency in strategy-proof object
allocation mechanisms. Journal of Economic Theory 170, 145-168

\item Nikzad, A. 2022. Rank-optimal assignments in uniform markets.
Theoretical Economics 17, 25--55.

\item Ortega, J., Klein, T. 2023. The cost of strategy-proofness in school
choice. Games and Economic Behavior 141, 515--528

\item Tasnim, M, Weesie, Y., Ghebreab, S., Baak, M. 2024. Strategic
manipulation of preferences in the rank minimization mechanism, Autonomous
Agents and Multi-Agent Systems 38(2), 44

\item Thomson, W. 2011. Fair Allocation Rules. Ch. 21 in Handbook of Social
Choice and Welfare, eds. by Arrow, K.-J., Sen, A., Suzumura, K., Vol. 2,
393--506.

\item Troyan, P. 2024. (Non-)obvious manipulability of rank-minimizing
mechanisms. Journal of Mathematical Economics 113, 103015

\item Troyan, P., Morrill T. 2020. Obvious manipulations. Journal of
Economic Theory 185, 104970
\end{description}

\newpage

\section*{Appendix}

\subsection*{Proof of Lemma 1}

First, by Remark 1, any $x^{\ast }\in \mathcal{X\setminus Y}$ can be
represented by a convex combination of deterministic assignments $%
y^{1},\cdots ,y^{I}$ with $I\geq 2$. Suppose not; that is, there is $y^{i}$
such that $RV\left( y^{i},p\right) >RV\left( x^{\ast },p\right) $. 
\begin{eqnarray}
RV\left( x^{\ast },p\right) &=&\sum\limits_{a\in \mathcal{A}%
}\sum\limits_{o\in \mathcal{O}}R\left( o,p_{a}\right) \times x_{ao}^{\ast } 
\notag \\
&=&\sum\limits_{a\in \mathcal{A}}\sum\limits_{o\in \mathcal{O}}\left(
R\left( o,p_{a}\right) \times \sum\limits_{i=1}^{I}\sigma
^{i}y_{ao}^{i}\right)  \notag \\
&=&\sum\limits_{i=1}^{I}\sigma ^{i}\sum\limits_{a\in \mathcal{A}%
}\sum\limits_{o\in \mathcal{O}}R\left( o,p_{a}\right) \times y_{ao}^{i} 
\notag \\
&=&\sum\limits_{i=1}^{I}\sigma ^{i}RV\left( y^{i},p\right) .  \label{b}
\end{eqnarray}%
Thus, $RV\left( y^{i},p\right) >RV\left( x^{\ast },p\right) $ implies that
there is $y^{j}$ such that $RV\left( y^{j},p\right) <RV\left( x^{\ast
},p\right) $, which contradicts that $x^{\ast }$ is a rank-minimizing
assignment.

\subsection*{Proof of Lemma 2}

Fix $p\in \mathcal{P}^{\left\vert \mathcal{A}\right\vert }$ and $a\in 
\mathcal{A}$. Suppose not; that is, for an assignment $x\in \mathcal{X}$
that is not wasteful for $p$, there is $o\in \mathcal{O}$ satisfying $%
R(o,p_{a})>\bar{k}\left( p_{a}\right) $ and $x_{ao}>0$. First, since $\sum_{%
\bar{o}\in \mathcal{O}}x_{a\bar{o}}=1$, 
\begin{equation*}
\sum\limits_{\bar{o}\in \mathcal{O}}\sum\limits_{\bar{a}\in \mathcal{A}}x_{%
\bar{a}\bar{o}}=\left\vert \mathcal{A}\right\vert \text{.}
\end{equation*}%
Moreover, since $x_{ao}>0$, 
\begin{eqnarray*}
\sum\limits_{\bar{o}:R(\bar{o},p_{a})\leq \bar{k}\left( p_{a}\right)
}\sum\limits_{\bar{a}\in \mathcal{A}}x_{\bar{a}o^{\prime }} &<&\left\vert 
\mathcal{A}\right\vert \leq \sum\limits_{\bar{o}:R(\bar{o},p_{a})\leq \bar{k}%
\left( p_{a}\right) }q_{o^{\prime }}\text{, and} \\
\sum\limits_{\bar{o}:R(\bar{o},p_{a})\leq \bar{k}\left( p_{a}\right) }x_{a%
\bar{o}} &<&1.
\end{eqnarray*}

Since $\sum_{\bar{a}\in \mathcal{A}}x_{\bar{a}o^{\prime \prime }}\leq
q_{o^{\prime \prime }}$ for all $o^{\prime \prime }\in \mathcal{O}$, there
is some $o^{\prime }\in \mathcal{O}$ such that $R(o^{\prime },p_{a})\leq 
\bar{k}\left( p_{a}\right) $ and $\sum_{\bar{a}\in \mathcal{A}}x_{\bar{a}%
o^{\prime }}<q_{o^{\prime }}$. Then, $\sum_{\bar{a}\in \mathcal{A}}x_{\bar{a}%
o^{\prime }}<q_{o^{\prime }},$ $x_{ao}>0,$ and $R(o^{\prime },p_{a})\leq 
\bar{k}\left( p_{a}\right) <R(o,p_{a})$ contradict that $x\in \mathcal{X}$
is not wasteful for $p$.

\subsection*{Proof of Proposition 1}

Let $p\in \mathcal{P}^{\left\vert \mathcal{A}\right\vert }$ satisfy $%
p_{a}^{k^{\prime }}=p_{a^{\prime }}^{k^{\prime }}$ for all $1,\cdots ,\bar{k}%
\left( p_{a}\right) $. Suppose not; that is, $\left( f^{U}\left( p\right)
\right) _{ao}>\left( f^{U}\left( p\right) \right) _{a^{\prime }o}$ for some $%
o\in \mathcal{O}$. Since $\left( f^{U}\left( p\right) \right) _{ao}>0$ and $%
f^{U}\left( p\right) $ is not wasteful, $R\left( o,p_{a}\right) \leq \bar{k}%
\left( p_{a}\right) $, by Lemma 2. Therefore, $R\left( o,p_{a}\right)
=R\left( o,p_{a^{\prime }}\right) $.

Now, let 
\begin{equation*}
Y=\left\{ y\in \mathcal{Y}^{\ast }\left( p\right) \left\vert \text{ }%
y_{ao}=1\right. \right\} ,Y^{\prime }=\left\{ y^{\ast }\in \mathcal{Y}^{\ast
}\left( p\right) \left\vert \text{ }y_{a^{\prime }o}^{\ast }=1\right.
\right\} .
\end{equation*}%
Since $\left( f^{U}\left( p\right) \right) _{ao}>0$, $Y\neq \emptyset $ and
thus let $Y=\left\{ y^{1},\cdots ,y^{\left\vert Y\right\vert }\right\} $.
Then, for each $i=1,\cdots ,\left\vert Y\right\vert $, there is $y^{i\prime
} $ be such that $y_{ao}^{i\prime }=0,$ $y_{a^{\prime }o}^{i\prime }=1,$ and 
$y_{a^{\prime \prime }o^{\prime }}^{i\prime }=y_{a^{\prime \prime }o^{\prime
}}^{i}$ for all $o^{\prime }\in \mathcal{O}\setminus \{o\}$ and all $%
a^{\prime \prime }\in \mathcal{A}\setminus \{a,a^{\prime }\}$. Since $%
R\left( o,p_{a}\right) =R\left( o,p_{a^{\prime }}\right) $, $RV\left(
y^{i},p\right) =RV\left( y^{i\prime },p\right) $ and therefore $y^{i\prime
}\in Y^{\prime }$. Moreover, since $y^{1\prime },$ $\cdots ,y^{\left\vert
Y\right\vert \prime }$ are distinct, $\left\vert Y\right\vert \leq
\left\vert Y^{\prime }\right\vert $. However, 
\begin{equation*}
\left( f^{U}\left( p\right) \right) _{ao}=\frac{\left\vert Y\right\vert }{%
\left\vert \mathcal{Y}^{\ast }\left( p\right) \right\vert }\leq \frac{%
\left\vert Y^{\prime }\right\vert }{\left\vert \mathcal{Y}^{\ast }\left(
p\right) \right\vert }=\left( f^{U}\left( p\right) \right) _{a^{\prime }o}
\end{equation*}%
contradicts $\left( f^{U}\left( p\right) \right) _{ao}>\left( f^{U}\left(
p\right) \right) _{a^{\prime }o}$.

\subsection*{Proof of Lemma 3}

Suppose $k\leq \bar{k}\left( p_{a}\right) $. Then, since $q_{\varnothing
}\geq \left\vert \mathcal{A}\right\vert $, $r_{a}^{k^{\prime
}}=p_{a}^{k^{\prime }}\neq \varnothing $ for all $k^{\prime }=1,\cdots ,k-1$.

When $k=1$, we let $R(\varnothing ,p_{a^{\prime }})=1$ for all $a^{\prime
}\in \mathcal{A\setminus }\left\{ a\right\} $. Then, 
\begin{equation*}
\left( f\left( r_{a},p_{-a}\right) \right) _{ar_{a}^{1}}=\left( f\left(
p_{a},p_{-a}\right) \right) _{ap_{a}^{1}}=1,
\end{equation*}%
where $R(r_{a}^{1},r_{a})=1<R(p_{a}^{1},r_{a})$. Thus, $a$ with $r_{a}$
strictly prefers $f\left( r_{a},p_{-a}\right) $ to $f\left(
p_{a},p_{-a}\right) $.

Next, we assume $k=2$. Then, $p_{a}^{1}=r_{a}^{1}$ and $p_{a}^{2}\neq
r_{a}^{2}$. We assume that under $\left( r_{a},p_{-a}\right) $, the agents
are divided into two groups. For the first group, there are $q_{r_{a}^{1}}+1$
agents whose preference is equivalent to $r_{a}$ under $\left(
r_{a},p_{-a}\right) $; that is, if $a^{\prime }$ belongs to this group, then 
$r_{a}=p_{a^{\prime }}$. Note that $a$ itself belongs to this group. For the
second group, the first-best object type is $\varnothing $ under $\left(
r_{a},p_{-a}\right) $. Then, since all agents in the second group always
assign $\varnothing $ with probability $1$ and $f$ satisfies ETE, 
\begin{eqnarray*}
\left( f\left( r_{a},p_{-a}\right) \right) _{ar_{a}^{1}} &=&\frac{%
q_{r_{a}^{1}}}{q_{r_{a}^{1}}+1}, \\
\left( f\left( r_{a},p_{-a}\right) \right) _{ar_{a}^{2}} &=&\frac{1}{%
q_{r_{a}^{1}}+1}.
\end{eqnarray*}%
On the other hand, 
\begin{eqnarray*}
\left( f\left( p_{a},p_{-a}\right) \right) _{ap_{a}^{1}} &=&\frac{%
q_{r_{a}^{1}}}{q_{r_{a}^{1}}+1}, \\
\left( f\left( p_{a},p_{-a}\right) \right) _{ap_{a}^{2}} &=&\frac{1}{%
q_{r_{a}^{1}}+1}.
\end{eqnarray*}%
Since $p_{a}^{1}=r_{a}^{1}$ and $p_{a}^{2}\neq r_{a}^{2}$, $a$ strictly
prefers $f\left( r_{a},p_{-a}\right) $ to $f\left( p_{a},p_{-a}\right) $.

We consider an arbitrary integer $k\geq 3$. Then, $r_{a}^{k^{\prime
}}=p_{a}^{k^{\prime }}$ for all $k^{\prime }=1,\cdots ,k-1$. We assume that
under $\left( r_{a},p_{-a}\right) $ the agents are divided into $k$ groups.
For the first group, there are $q_{r_{a}^{1}}+1$ agents whose preference is
equivalent to $r_{a}$ under $\left( r_{a},p_{-a}\right) $; that is, if $%
a^{\prime }$ is in this group, then $r_{a}=p_{a^{\prime }}$. Note that $a$
belongs to this group. Next, we consider $j$th group for each $j=2,3,\cdots
,k-1$. In $j$th group, there are just $q_{r_{a}^{j}}$ agents whose
first-best object type is $r_{a}^{j}$ and the second-best object is $%
r_{a}^{j+1},$ the third-best $r_{a}^{j+2},$ $\cdots $ and the $k$th-best $%
r_{a}^{k}$ under $\left( r_{a},p_{-a}\right) $. If $a_{j}$ belongs $j$th
group for $j=1,2,\cdots ,k-1$, the preference orders of them are as follows:%
\begin{eqnarray}
p_{a_{1}} &=&\left( r_{a}^{1},r_{a}^{2},\ldots
,r_{a}^{k-2},r_{a}^{k-1},r_{a}^{k},\ldots \right) ,  \label{c} \\
p_{a_{2}} &=&\left( r_{a}^{2},r_{a}^{3},\ldots
,r_{a}^{k-1},r_{a}^{1},r_{a}^{k},,\ldots \right) ,  \notag \\
&&\vdots  \notag \\
p_{a_{j}} &=&\left( r_{a}^{j},r_{a}^{j+1},\ldots
,r_{a}^{j-2},r_{a}^{j-1},r_{a}^{k},\ldots \right) ,  \notag \\
&&\vdots  \notag \\
p_{a_{k-1}} &=&\left( r_{a}^{k-1},r_{a}^{1},\ldots
,r_{a}^{k-3},r_{a}^{k-2},r_{a}^{k},\ldots \right) .  \notag
\end{eqnarray}%
Since $k\leq \bar{k}\left( p_{a}\right) $, 
\begin{equation*}
q_{r_{a}^{1}}+q_{r_{a}^{2}}+\cdots
+q_{r_{a}^{k-1}}=\sum\limits_{o:R(o,r_{a})<k}q_{o}<\left\vert \mathcal{A}%
\right\vert .
\end{equation*}%
and thus $\left( r_{a},p_{-a}\right) $ is well-defined so far. Finally, for
all the remaining agents (belonging to the $k$th group), the first-best
object type is $\varnothing $ under $\left( r_{a},p_{-a}\right) $.

In this case, at least one agent cannot be assigned to their first-best
object. Thus, in a rank-minimizing assignment, all agents in $j$th group for
all $j=2,3,\cdots ,k-1$ is assigned to their first-best object type $%
r_{a}^{j}$, and $q_{r_{a}^{1}}$ agents in the first group is also assigned
to their first-best object type $r_{a}^{1},$ and moreover, just one agent in
the first group is also assigned to $r_{a}^{k}$. Moreover, an agent who
belongs to the $k$th group is assigned to their first-best object type $%
\varnothing $.

Since all agents excepting those belonging to the first group are assigned
to their first-best alternative with probability $1$ and $f^{U}$ satisfies
ETE, 
\begin{eqnarray*}
\left( f\left( r_{a},p_{-a}\right) \right) _{ar_{a}^{1}} &=&\frac{%
q_{r_{a}^{1}}}{q_{r_{a}^{1}}+1}, \\
\left( f\left( r_{a},p_{-a}\right) \right) _{ar_{a}^{k}} &=&\frac{1}{%
q_{r_{a}^{1}}+1}.
\end{eqnarray*}%
On the other hand, 
\begin{eqnarray*}
\left( f\left( p_{a},p_{-a}\right) \right) _{ap_{a}^{1}} &=&\frac{%
q_{r_{a}^{1}}}{q_{r_{a}^{1}}+1}, \\
\left( f\left( p_{a},p_{-a}\right) \right) _{ap_{a}^{k}} &=&\frac{1}{%
q_{r_{a}^{1}}+1}.
\end{eqnarray*}%
Since $p_{a}^{k^{\prime }}=r_{a}^{k^{\prime }}$ for all $k^{\prime
}=1,\cdots ,k-1$ and $p_{a}^{k}\neq r_{a}^{k}$, $a$ strictly prefers $%
f^{U}\left( r_{a},p_{-a}\right) $ to $f^{U}\left( p_{a},p_{-a}\right) $.
Thus, we have the first result.

\subsection*{Proof of Proposition 2}

Let $f$ be a fair mechanism, $p_{a}\neq r_{a}$ and $k$ be the smallest
integer such that $p_{ak}\neq r_{ak}$. First, suppose $k\leq \bar{k}\left(
p_{a}\right) $. By Lemma 2, there is $p_{-a}\in \mathcal{P}^{\left\vert 
\mathcal{A}\right\vert -1}$ such that $a$ strictly prefers $f\left(
r_{a},p_{-a}\right) $ to $f\left( p_{a},p_{-a}\right) $. Therefore, in this
case, $r_{a}$ is not dominate by $p_{a}$.

Second, suppose $k>\bar{k}\left( p_{a}\right) $. In this case, since $f$
satisfies ETE, $f\left( r_{a},p_{-a}\right) =f\left( p_{a},p_{-a}\right) $.
Therefore, $p_{a}$ does not dominate $r_{a}$\textbf{.}

\subsection*{Proof of Theorem 1}

Fix any $\left( r_{a},p_{-a}\right) \in \mathcal{P}^{\left\vert \mathcal{A}%
\right\vert }$ and any $d_{a}\in \mathcal{D}_{a}$. We show that $a$ weakly
prefers $g\left( f^{U}\left( d_{a},p_{-a}\right) ,r\right) $ to $g\left(
f^{U}\left( r_{a},p_{-a}\right) ,r\right) $. First, suppose that there is no 
$y\in \mathcal{Y}^{\ast }\left( r_{a},p_{-a}\right) $ such that $y_{ao}=1$
for some $o\in \mathcal{O}$ with $R\left( o,r_{a}\right) <R\left(
\varnothing ,r_{a}\right) $. Then, $g\left( f^{U}\left( r_{a},p_{-a}\right)
,r\right) _{a\varnothing }=1.$ Thus, for any assignment $x$, $a$ weakly
prefers $g\left( x,r\right) $ to $g\left( f^{U}\left( r_{a},p_{-a}\right)
,r\right) $.

Second, suppose that there is some $y\in \mathcal{Y}^{\ast }\left(
r_{a},p_{-a}\right) $ such that $y_{ao}=1$ with $R\left( o,r_{a}\right)
<R\left( \varnothing ,r_{a}\right) $. In this case, we have the following
two results.

\begin{claim}
If $y\in \mathcal{Y}^{\ast }\left( r_{a},p_{-a}\right) $ such that $y_{ao}=1$
where $R\left( o,r_{a}\right) <R\left( \varnothing ,r_{a}\right) $, then $%
y\in \mathcal{Y}^{\ast }\left( d_{a},p_{-a}\right) $.
\end{claim}

\textbf{Proof.} By the construction, 
\begin{equation*}
RV\left( y,\left( r_{a},p_{-a}\right) \right) =RV\left( y,\left(
d_{a},p_{-a}\right) \right) .
\end{equation*}

Suppose $y\in \mathcal{Y}^{\ast }\left( r_{a},p_{-a}\right) $ such that $%
y_{ao}=1$ where $R\left( o,r_{a}\right) <R\left( \varnothing ,r_{a}\right) $%
. Toward a contradiction, suppose $y\notin \mathcal{Y}^{\ast }\left(
d_{a},p_{-a}\right) $.\ 

We first consider the case where there is $y^{\prime }\in \mathcal{Y}^{\ast
}\left( d_{a},p_{-a}\right) $ that satisfies $y_{ao^{\prime }}^{\prime }=1$
for $o^{\prime }$ such that $R\left( o^{\prime },r_{a}\right) <R\left(
\varnothing ,r_{a}\right) $. By the construction of $d_{a}$,%
\begin{eqnarray*}
RV\left( y^{\prime },\left( r_{a},p_{-a}\right) \right) &=&RV\left(
y^{\prime },\left( d_{a},p_{-a}\right) \right) < \\
RV\left( y,\left( d_{a},p_{-a}\right) \right) &=&RV\left( y,\left(
r_{a},p_{-a}\right) \right) ,
\end{eqnarray*}%
which contradicts $y\in \mathcal{Y}^{\ast }\left( r_{a},p_{-a}\right) $.

We second consider the case where there is no such a deterministic
assignment. Since $\mathcal{Y}^{\ast }\left( d_{a},p_{-a}\right) \neq
\emptyset $ (by Corollary 1), there is $y^{\prime }\in \mathcal{Y}^{\ast
}\left( d_{a},p_{-a}\right) $ such that $y_{ao^{\prime }}^{\prime }=1$ for $%
o $ such that $R\left( o,r_{a}\right) \geq R\left( \varnothing ,r_{a}\right) 
$. In this case, 
\begin{equation*}
RV\left( y,\left( r_{a},p_{-a}\right) \right) =RV\left( y,\left(
d_{a},p_{-a}\right) \right) >RV\left( y^{\prime },\left( d_{a},p_{-a}\right)
\right) .
\end{equation*}%
Now, we consider $y^{\prime \prime }$ such that $y_{a\varnothing }^{\prime
\prime }=1,$ $y_{a\bar{o}}^{\prime \prime }=0$ for all $\bar{o}\in \mathcal{O%
}\setminus \left\{ \varnothing \right\} $ and $y_{a^{\prime }\bar{o}%
}^{\prime \prime }=y_{a^{\prime }\bar{o}}$ for all $a^{\prime }\in \mathcal{A%
}\setminus \left\{ a\right\} $ and all $\bar{o}\in \mathcal{O}$. Since $%
q_{\varnothing }\geq \left\vert \mathcal{A}\right\vert $, $y^{\prime \prime
} $ is an assignment. Moreover, since $R\left( o,r_{a}\right) \geq R\left(
\varnothing ,r_{a}\right) $, 
\begin{equation*}
RV\left( y^{\prime },\left( d_{a},p_{-a}\right) \right) \geq RV\left(
y^{\prime \prime },\left( r_{a},p_{-a}\right) \right) \text{.}
\end{equation*}%
Then, 
\begin{equation*}
RV\left( y,\left( r_{a},p_{-a}\right) \right) >RV\left( y^{\prime \prime
},\left( r_{a},p_{-a}\right) \right)
\end{equation*}%
contradicts $y\in \mathcal{Y}^{\ast }\left( r_{a},p_{-a}\right) $. \textbf{%
Q.E.D.}\newline

\begin{claim}
Suppose that there is $y^{\ast }\in \mathcal{Y}^{\ast }\left(
r_{a},p_{-a}\right) $ such that $y_{ao}^{\ast }=1$ and $R\left(
o,r_{a}\right) <R\left( \varnothing ,r_{a}\right) $. Then, $\left\vert 
\mathcal{Y}^{\ast }\left( d_{a},p_{-a}\right) \right\vert \leq \left\vert 
\mathcal{Y}^{\ast }\left( r_{a},p_{-a}\right) \right\vert $.
\end{claim}

\textbf{Proof.} Suppose that there is $y^{\ast }\in \mathcal{Y}^{\ast
}\left( r_{a},p_{-a}\right) $ such that $y_{ao}^{\ast }=1$ and $R\left(
o,r_{a}\right) <R\left( \varnothing ,r_{a}\right) $. Then, by Claim 1, $%
y^{\ast }\in \mathcal{Y}^{\ast }\left( d_{a},p_{-a}\right) $.

Next, we show that if $y\in \mathcal{Y}^{\ast }\left( d_{a},p_{-a}\right) $,
then either $y\in \mathcal{Y}^{\ast }\left( r_{a},p_{-a}\right) $ or $%
y^{\prime }\in \mathcal{Y}^{\ast }\left( r_{a},p_{-a}\right) ,$ such that $%
y_{a\varnothing }^{\prime }=1$ and $y_{a^{\prime }o}^{\prime }=y_{a^{\prime
}o}$ for all $a^{\prime }\in \mathcal{A}\setminus \left\{ a\right\} $ and $%
o\in \mathcal{O}$. Since $q_{\varnothing }\geq \left\vert \mathcal{A}%
\right\vert $, $y^{\prime }\in \mathcal{Y}$.

Suppose that $y\in \mathcal{Y}^{\ast }\left( d_{a},p_{-a}\right) $ and $%
y\notin \mathcal{Y}^{\ast }\left( r_{a},p_{-a}\right) $. We have%
\begin{equation*}
RV\left( y,\left( d_{a},p_{-a}\right) \right) =RV\left( y^{\ast },\left(
d_{a},p_{-a}\right) \right) =RV\left( y^{\ast },\left( r_{a},p_{-a}\right)
\right) <RV\left( y,\left( r_{a},p_{-a}\right) \right) .
\end{equation*}%
Then, by the construction of $d_{a}$, $y_{ao^{\prime }}=1\,$such that $%
R\left( o^{\prime },r_{a}\right) \geq R\left( \varnothing ,r_{a}\right) $
and $R\left( o^{\prime },d_{a}\right) \geq R\left( \varnothing ,r_{a}\right) 
$. Then, we have 
\begin{equation*}
RV\left( y^{\prime },\left( r_{a},p_{-a}\right) \right) \leq RV\left(
y,\left( d_{a},p_{-a}\right) \right) \leq RV\left( y^{\ast },\left(
d_{a},p_{-a}\right) \right) \text{.}
\end{equation*}%
Since $y^{\ast }\in \mathcal{Y}^{\ast }\left( r_{a},p_{-a}\right) $, $%
y^{\prime }\in \mathcal{Y}^{\ast }\left( r_{a},p_{-a}\right) $. Therefore,
if $y\in \mathcal{Y}^{\ast }\left( d_{a},p_{-a}\right) $, then either $y\in 
\mathcal{Y}^{\ast }\left( r_{a},p_{-a}\right) $ or $y^{\prime }\in \mathcal{Y%
}^{\ast }\left( r_{a},p_{-a}\right) $ defined above.

Finally, we assume $\left\vert \mathcal{Y}^{\ast }\left( d_{a},p_{-a}\right)
\right\vert \geq 2$ and arbitrarily let $y_{1},y_{2}\in \mathcal{Y}^{\ast
}\left( d_{a},p_{-a}\right) $ such that $y_{1}\neq y_{2}$. For $i=1,2$, let $%
y_{i}^{\prime }$ be such that $\left( y_{i}^{\prime }\right) _{a\varnothing
}=1$ and $\left( y_{i}^{\prime }\right) _{a^{\prime }o}=\left( y_{i}\right)
_{a^{\prime }o}$ for all $a^{\prime }\in \mathcal{A}\setminus \left\{
a\right\} $ and $o\in \mathcal{O}$.

We show $y_{1}^{\prime }\neq y_{2}^{\prime }$. Suppose not; that is, $%
y_{1}^{\prime }=y_{2}^{\prime }$. Then, $\left( y_{1}\right) _{a^{\prime
}o}=\left( y_{2}\right) _{a^{\prime }o}$ for all $a^{\prime }\in \mathcal{A}%
\setminus \left\{ a\right\} $ and $o\in \mathcal{O}$. Since $y_{1}\neq y_{2}$%
, we have $o_{1}\neq o_{2}$ where $\left( y_{1}\right) _{ao_{1}}=1$ and $%
\left( y_{2}\right) _{ao_{2}}=1$. Without loss of generality, we assume $%
R\left( o_{1},d_{a}\right) <R\left( o_{2},d_{a}\right) $. However, $R\left(
o_{1},d_{a}\right) <R\left( o_{2},d_{a}\right) $ and $\left( y_{1}\right)
_{a^{\prime }o}=\left( y_{2}\right) _{a^{\prime }o}$ for all $a^{\prime }\in 
\mathcal{A}\setminus \left\{ a\right\} $ and $o\in \mathcal{O}$ imply 
\begin{equation*}
RV\left( y_{1},\left( d_{a},p_{-a}\right) \right) <RV\left( y_{2},\left(
d_{a},p_{-a}\right) \right)
\end{equation*}%
contradicting that $y_{2}\in \mathcal{Y}^{\ast }\left( d_{a},p_{-a}\right) $%
. Therefore, $y_{1}^{\prime }\neq y_{2}^{\prime }$.

By these results, if $\left\{ y_{1},y_{2},\cdots ,y_{n}\right\} =\mathcal{Y}%
^{\ast }\left( d_{a},p_{-a}\right) ,$ then $\left\{ y_{1}^{\prime \prime
},y_{2}^{\prime \prime },\cdots ,y_{n}^{\prime \prime }\right\} \subseteq 
\mathcal{Y}^{\ast }\left( r_{a},p_{-a}\right) $ where $y_{i}^{\prime \prime
} $ is either $y_{i}$ or $y_{i}^{\prime }$ constructed above. Since $%
y_{1}^{\prime },y_{2}^{\prime },\cdots ,y_{n}^{\prime }$ are distinct, $%
y_{1}^{\prime \prime },y_{2}^{\prime \prime },\cdots ,y_{n}^{\prime \prime }$
are also distinct and thus $\left\vert \mathcal{Y}^{\ast }\left(
d_{a},p_{-a}\right) \right\vert \leq \left\vert \mathcal{Y}^{\ast }\left(
r_{a},p_{-a}\right) \right\vert $. \textbf{Q.E.D. }\newline

Now, we show that $a$ weakly prefers $g\left( f^{U}\left(
d_{a},p_{-a}\right) ,r\right) $ to $g\left( f^{U}\left( r_{a},p_{-a}\right)
,r\right) $. By Claim 2, 
\begin{equation*}
\left\vert \mathcal{Y}^{\ast }\left( d_{a},p_{-a}\right) \right\vert \leq
\left\vert \mathcal{Y}^{\ast }\left( r_{a},p_{-a}\right) \right\vert \text{.}
\end{equation*}%
Moreover, by Claim 1, any $y^{\ast }\in \mathcal{Y}^{\ast }\left(
r_{a},p_{-a}\right) $ such that $y_{ao}^{\ast }=1$ with $R\left(
o,r_{a}\right) <R\left( \varnothing ,r_{a}\right) $ is also an element of $%
\mathcal{Y}^{\ast }\left( d_{a},p_{-a}\right) $.

Thus, for any $o\in \mathcal{O}$ such that $R\left( o,r_{a}\right) <R\left(
\varnothing ,r_{a}\right) $,%
\begin{eqnarray*}
\left( g\left( f^{U}\left( d_{a},p_{-a}\right) ,r\right) \right) _{ao}
&=&\left( f^{U}\left( d_{a},p_{-a}\right) \right) _{ao}\geq \\
\left( f^{U}\left( r_{a},p_{-a}\right) \right) _{ao} &=&\left( g\left(
f^{U}\left( r_{a},p_{-a}\right) ,r\right) \right) _{ao}\text{.}
\end{eqnarray*}%
Since $g\left( f^{U}\left( d_{a},p_{-a}\right) \right) _{ao^{\prime }}=0$
for all $R\left( o^{\prime },r_{a}\right) >R\left( \varnothing ,r_{a}\right) 
$, 
\begin{equation*}
\left( g\left( f^{U}\left( d_{a},p_{-a}\right) ,r\right) \right)
_{a\varnothing }\leq \left( g\left( f^{U}\left( r_{a},p_{-a}\right)
,r\right) \right) _{a\varnothing }.
\end{equation*}%
Thus, $a$ weakly prefers $g\left( f\left( d_{a},p_{-a}\right) ,r\right) $ to 
$g\left( f\left( r_{a},p_{-a}\right) ,r\right) $.

\subsection*{Proof of Theorem 2}

Suppose that there are $a\in \mathcal{A}$, $o,o^{\prime }\in \mathcal{O}$
satisfying (\ref{f}) and let$\ d_{a}\in \mathcal{D}_{a}$ such that $R\left(
\varnothing ,r_{a}\right) =R\left( o^{\prime },d_{a}\right) $. By Theorem 1,
it is sufficient to show that there is $p_{-a}\in \mathcal{P}^{\left\vert 
\mathcal{A}\right\vert -1}$ such that $a$ strictly prefers $g\left(
f^{U}\left( d_{a},p_{-a}\right) ,r\right) $ to $g\left( f^{U}\left(
r_{a},p_{-a}\right) ,r\right) $. Let $p_{-a}$ be such that $p_{a^{\prime
}}=d_{a}$ for all $a^{\prime }\in \mathcal{A\setminus }\left\{ a\right\} $.

First, we show that for all $y\in \mathcal{Y}^{\ast }\left(
r_{a},p_{-a}\right) $, $y_{a\varnothing }=0$. Suppose not; that is, there is 
$y\in \mathcal{Y}^{\ast }\left( r_{a},p_{-a}\right) $, $y_{a\hat{o}}=1$
where $R\left( \hat{o},r_{a}\right) <R\left( \varnothing ,r_{a}\right) $.
Then, by (\ref{f}), there is $a^{\prime }\in \mathcal{A}$ such that $%
y_{a^{\prime }o^{\prime \prime }}=1$ with $R\left( o^{\prime \prime
},p_{a^{\prime }}\right) >R\left( o^{\prime },p_{a^{\prime }}\right) \left(
=R\left( \varnothing ,r_{a}\right) \right) ,$ because $p_{a^{\prime }}=d_{a}$
for all $a^{\prime }\in \mathcal{A\setminus }\left\{ a\right\} $. Now, let $%
y^{\prime }\in \mathcal{Y}$ such that $y_{a^{\prime }\hat{o}}^{\prime }=1,$ $%
y_{a\varnothing }^{\prime }=1,$ and $y_{a^{\prime \prime }\bar{o}}^{\prime
}=y_{a^{\prime \prime }\bar{o}}$ for all $a^{\prime \prime }\in \mathcal{%
A\setminus }\left\{ a,a^{\prime }\right\} $ and all $\bar{o}\in \mathcal{O}$%
. Then, since%
\begin{gather*}
R\left( \hat{o},r_{a^{\prime }}\right) =R\left( \hat{o},p_{a^{\prime
}}\right) \text{ and }R\left( o^{\prime \prime },p_{a^{\prime }}\right)
>R\left( \varnothing ,r_{a}\right) , \\
RV\left( y^{\prime },\left( r_{a},p_{-a}\right) \right) <RV\left( y,\left(
r_{a},p_{-a}\right) \right) ,
\end{gather*}
contradicting $y\in \mathcal{Y}^{\ast }\left( r_{a},p_{-a}\right) $.\
Therefore, for all $y\in \mathcal{Y}^{\ast }\left( r_{a},p_{-a}\right) $, $%
y_{a\varnothing }=0$.

By this result, 
\begin{equation*}
\left( f^{U}\left( r_{a},p_{-a}\right) ,r\right) _{a\varnothing }=\left(
g\left( f^{U}\left( d_{a},p_{-a}\right) ,r\right) \right) _{a\varnothing }=1
\end{equation*}%
On the other hand, since $f^{U}$ satisfies ETE,%
\begin{equation*}
\left( f^{U}\left( d_{a},p_{-a}\right) ,r\right) _{a\hat{o}}=\left( g\left(
f^{U}\left( d_{a},p_{-a}\right) ,r\right) \right) _{a\hat{o}}>0.
\end{equation*}%
Therefore, $a$ strictly prefers $g\left( f^{U}\left( d_{a},p_{-a}\right)
,r\right) $ to $g\left( f^{U}\left( r_{a},p_{-a}\right) ,r\right) $.

\subsection*{Proof of Proposition 3}

Let $r_{a}=\left( o_{1},\varnothing ,\cdots \right) $ and $r_{a^{\prime
}}=\left( o_{1},o_{2},\varnothing ,\cdots \right) $ for all $a^{\prime }\in 
\mathcal{A\setminus }\left\{ a\right\} $. Furthermore, let $d=\left( d_{\bar{%
a}}\right) _{\bar{a}\in \mathcal{A}}$ such that $d_{\bar{a}%
}=(o_{1},o_{2},o_{3},\cdots ,\varnothing )$ for all $\bar{a}\in \mathcal{A}$%
; that is, $d$ is a profile in which every agent adopts an ODS.

First, we show that no agent has an incentive to unilaterally deviate from $%
d $. It is sufficient to show that for all $\bar{a}\in \mathcal{A}$ and all $%
p_{\bar{a}}\in \mathcal{P}$, either 
\begin{equation}
\left( f^{U}\left( d\right) \right) _{\bar{a}o_{1}}>\left( f^{U}\left( p_{%
\bar{a}},d_{_{-\bar{a}}}\right) \right) _{\bar{a}o_{1}}  \label{z1}
\end{equation}%
or 
\begin{equation}
\left( f^{U}\left( d\right) \right) _{\bar{a}\bar{o}}=\left( f^{U}\left( p_{%
\bar{a}},d_{_{-\bar{a}}}\right) \right) _{\bar{a}\bar{o}},\,\ \text{for all }%
\bar{o}\in \mathcal{O},  \label{z2}
\end{equation}%
because $o$ is the best object type for all agents.

Let $k$ be the smallest integer such that $p_{\bar{a}k}\neq d_{\bar{a}k}$.
By Lemma 2, if $k>\bar{k}\left( p_{\bar{a}}\right) $, then (\ref{z2}) is
satisfied. We assume $k\leq \bar{k}\left( p_{_{\bar{a}}}\right) $; that is,%
\begin{equation*}
\sum\limits_{\bar{o}:R(\bar{o},p_{\bar{a}})<k}q_{\bar{o}}\left(
=\sum\limits_{\bar{o}:R(\bar{o},d_{\bar{a}})<k}q_{\bar{o}}\right)
<\left\vert \mathcal{A}\right\vert
\end{equation*}%
is satisfied for that $k$. Let $o^{\prime \prime }$ such that $R\left(
o^{\prime \prime },p_{\bar{a}}\right) =k<R\left( o^{\prime \prime },d_{\bar{a%
}}\right) $. Let $y\in \mathcal{Y}^{\ast }\left( d\right) $ such that $y_{%
\bar{a}o_{k}}=1$. Moreover, we arbitrarily choose $y^{\prime }$ such that $%
y_{\bar{a}o_{1}}^{\prime }=1$.

We consider two cases. First, we assume 
\begin{equation*}
\sum\limits_{\bar{o}:R(\bar{o},d_{\bar{a}})\leq k}q_{\bar{o}}=\sum\limits_{%
\bar{o}:R(\bar{o},d_{\bar{a}})<k}q_{\bar{o}}+q_{o_{k}}<\left\vert \mathcal{A}%
\right\vert .
\end{equation*}%
In this case, some agent is assigned to $o_{k+1}$ in a rank-minimizing
assignment under $d$. Let $y^{\prime \prime }$ be such that $y_{\bar{a}%
o^{\prime \prime }}^{\prime \prime }=1$, $y_{a^{\prime }o_{k}}^{\prime
\prime }=1$ for $a^{\prime }$ with $y_{a^{\prime }o_{k+1}}=1,$ and $%
y_{a^{\prime \prime }\bar{o}}^{\prime \prime }=y_{a^{\prime \prime }\bar{o}}$
for all $a^{\prime \prime }\in \mathcal{A\setminus }\left\{ \bar{a}%
,a^{\prime }\right\} $ and $\bar{o}\in \mathcal{O}$. Then, 
\begin{equation*}
RV(y^{\prime \prime },\left( p_{\bar{a}},d_{-\bar{a}}\right) )<RV(y,d)\leq
RV(y^{\prime },d)\leq RV(y^{\prime },\left( p_{\bar{a}},d_{-\bar{a}}\right)
).
\end{equation*}%
Thus, $y^{\prime }\notin \mathcal{Y}^{\ast }\left( p_{\bar{a}},d_{-\bar{a}%
}\right) $ and (\ref{z1}) is satisfied in this case.

Second, we assume 
\begin{equation*}
\sum\limits_{\bar{o}:R(\bar{o},d_{\bar{a}})\leq k}q_{\bar{o}}=\sum\limits_{%
\bar{o}:R(\bar{o},d_{\bar{a}})<k}q_{\bar{o}}+q_{o_{k}}\geq \left\vert 
\mathcal{A}\right\vert .
\end{equation*}%
Then, since $q_{o_{1}}<\left\vert \mathcal{A}\right\vert $, $k>1$. Let $%
y^{\prime \prime }$ be such that $y_{\bar{a}o^{\prime \prime }}^{\prime
\prime }=1$, and $y_{a^{\prime \prime }\bar{o}}^{\prime \prime
}=y_{a^{\prime \prime }\bar{o}}$ for all $a^{\prime \prime }\in \mathcal{%
A\setminus }\left\{ \bar{a}\right\} $ and $\bar{o}\in \mathcal{O}$.
Moreover, let $y^{\prime \prime \prime }$ be an arbitrary assignment in $%
\mathcal{Y}^{\ast }\left( d\right) $ such that $y_{\bar{a}o_{k^{\prime
}}}^{\prime \prime \prime }=1$ for some $k^{\prime }=1,\cdots ,k-1$.%
\begin{equation*}
RV(y^{\prime \prime },\left( p_{\bar{a}},d_{-\bar{a}}\right)
)=RV(y,d)=RV(y^{\prime \prime \prime },d)=RV(y^{\prime \prime \prime
},\left( p_{\bar{a}},d_{-\bar{a}}\right) ).
\end{equation*}%
Therefore, (\ref{z2}) is satisfied in this case.

Finally, suppose $q_{o_{1}}+q_{o_{2}}<\left\vert \mathcal{A}\right\vert $.
We show that $g\left( f^{U}\left( d\right) ,r\right) $ is wasteful. Since $%
q_{o_{1}}+q_{o_{2}}<\left\vert \mathcal{A}\right\vert $, there is $y\in 
\mathcal{Y}^{\ast }\left( d\right) $, $y_{ao_{2}}=1$ and $y_{a^{\prime
}o_{1}}=y_{a^{\prime }o_{2}}=0$ for some $a^{\prime }\in \mathcal{A\setminus 
}\left\{ a\right\} $. Thus, $\left( f^{U}\left( d\right) \right) _{ao_{2}}>0$
and $\left( g\left( f^{U}\left( d\right) ,r\right) \right) _{ao_{2}}=0$.
Therefore, $g\left( f^{U}\left( d\right) ,r\right) $ is wasteful.

\subsection*{Proof of Proposition 4}

First, suppose that $p$ satisfies all of Conditions\ (i), (ii) and (iii).
Then, for any two agents $a_{1}$ and $a_{2}$ belonging to different groups, $%
p_{a_{1}}$ and $p_{a_{2}}$ are not essentially the same. Moreover, for any
two agents $a_{1}$ and $a_{2}$ belonging to the same group (either the
second group or the third group), $\left( f^{M}\left( p\right) \right)
_{a_{1}o^{\prime }}=\left( f^{M}\left( p\right) \right) _{a_{2}o^{\prime }}$
for all $o^{\prime }\in \mathcal{O}$.

Second if $p$ does not satisfies all of Conditions\ (i), (ii) and (iii),
then $f^{M}\left( p\right) =f^{U}\left( p\right) $. By Proposition 2, $f^{M}$
satisfies ETE.

\subsection*{Proof of Proposition 5}

First, if $r_{a}^{1}=\varnothing ,$ then $\left( f^{M}\left(
r_{a},p_{-a}\right) \right) _{a\varnothing }=1$ for all $p_{-a}\in \mathcal{P%
}^{\left\vert \mathcal{A}\right\vert -1}$. Thus, in this case, there is no $%
p_{a}\in \mathcal{P}$ that dominate $r_{a}$. Thus, we assume $r_{a}^{1}\neq
\varnothing $. We consider $p_{a}\left( \neq r_{a}\right) $. Let $k$ be the
smallest integer such that $p_{a}^{k}\neq r_{a}^{k}$.

First, suppose $k\leq \bar{k}\left( p_{a}\right) $ and $r_{a}^{k}\neq
\varnothing $. Since $k\leq \bar{k}\left( p_{a}\right) $, $r_{a}^{k^{\prime
}}\neq \varnothing $ for all $k^{\prime }=1,\cdots ,k-1$; that is, $%
R(\varnothing ,r_{a})>k$ and $R(\varnothing ,p_{a})\geq k$. In this case,
since $f^{M}$ satisfies ETE (by Proposition 4), we revisit the examples in
the first result of Lemma 3. When $k=1$ or $2$, trivially, we can use the
example as is, because 
\begin{eqnarray*}
\left( g\left( f^{M}\left( r_{a},p_{-a}\right) ,r\right) \right) _{ao}
&=&\left( f^{M}\left( r_{a},p_{-a}\right) ,r\right) _{ao}, \\
\left( g\left( f^{M}\left( p_{a},p_{-a}\right) ,r\right) \right) _{ao}
&=&\left( f^{M}\left( p_{a},p_{-a}\right) ,r\right) _{ao}.
\end{eqnarray*}%
for all $o\in \mathcal{O}$.

We consider $k=3$. Then, since $R(\varnothing ,r_{a})>k$ and $R(\varnothing
,p_{a})\geq k$, 
\begin{eqnarray*}
\left( g\left( f^{M}\left( r_{a},p_{-a}\right) ,r\right) \right)
_{ar_{a}^{1}} &=&\frac{q_{r_{a}^{1}}}{q_{r_{a}^{1}}+1},\text{ }\left(
g\left( f^{M}\left( r_{a},p_{-a}\right) ,r\right) \right) _{ar_{a}^{k}}=%
\frac{1}{q_{r_{a}^{1}}+1}. \\
\left( g\left( f^{M}\left( r_{a},p_{-a}\right) ,r\right) \right)
_{ap_{a}^{1}} &=&\frac{q_{r_{a}^{1}}}{q_{r_{a}^{1}}+1},
\end{eqnarray*}%
and moreover, either%
\begin{eqnarray*}
\left( g\left( f^{M}\left( p_{a},p_{-a}\right) ,r\right) \right)
_{ap_{a}^{k}} &=&\frac{1}{q_{r_{a}^{1}}+1}\text{ or} \\
\left( g\left( f^{M}\left( p_{a},p_{-a}\right) ,r\right) \right)
_{a\varnothing } &=&\frac{1}{q_{r_{a}^{1}}+1}\text{.}
\end{eqnarray*}%
Since $R(\varnothing ,r_{a})>k$ and $R(p_{a}^{k},r_{a})>k$, in either case, $%
a$ prefers $g\left( f^{M}\left( r_{a},p_{-a}\right) ,r\right) $ to $g\left(
f^{M}\left( p_{a},p_{-a}\right) ,r\right) $.

Second, suppose $k\leq \bar{k}\left( p_{a}\right) $ and $r_{a}^{k}=%
\varnothing $. Then, since we assume $r_{a}^{1}\neq \varnothing $, $k\geq 2$%
. Moreover, in this case, $p_{a}^{k^{\prime }}\neq \varnothing $ for all $%
k^{\prime }=1,\cdots ,k$. Therefore, $R(\varnothing ,p_{a})>R(\varnothing
,r_{a})=k$. Let $p_{-a}$ be such that there are $q_{o}$ students represented
by $a^{\prime }$ whose revealed preference is equivalent to $%
r_{a}(=p_{a^{\prime }})$ and the first-best type of the other agents
represented by $a^{\prime \prime }$ is $\varnothing $. Then, $\left(
p_{a},p_{-a}\right) $ satisfies (i), (ii) and (iii) with $l=k$, because $%
k\leq \bar{k}\left( p_{a}\right) $. Moreover, $\left( r_{a},p_{-a}\right) $
does not satisfy all of them. Therefore, in this case, $g\left( f^{M}\left(
r_{a},p_{-a}\right) \right) $ is more preferable for $a$ than $g\left(
f^{M}\left( p_{a},p_{-a}\right) \right) $.

Third, suppose $k>\bar{k}\left( p_{a}\right) $. In this case, $r_{a}$ and $%
p_{a}$ are essentially the same. Since $f^{M}$ satisfies ETE (by Proposition
5), 
\begin{equation*}
g\left( f^{M}\left( r_{a},p_{-a}\right) ,r\right) =g\left( f^{M}\left(
p_{a},p_{-a}\right) ,r\right)
\end{equation*}%
for all $p_{-a}\in \mathcal{P}^{\left\vert \mathcal{A}\right\vert -1}$.

Therefore, for any $r_{a}\in \mathcal{P}$, there is no $p_{a}\in \mathcal{P}$
that dominates $r_{a}$.

\subsection*{Proof of Proposition 6}

Suppose not; that is, there exist $(r_{a},p_{-a})\in \mathcal{P}^{\left\vert 
\mathcal{A}\right\vert }$ such that $a$ strictly prefers $g\left(
f^{U}\left( \pi _{a},p_{-a}\right) ,r\right) $ to $g\left( f^{U}\left(
r_{a},p_{-a}\right) ,r\right) $ where $\pi _{a}$ is an OPS of $a$. By the
definition of OPS, when $o$ and $o^{\prime }$ satisfy $R\left(
o,r_{a}\right) <R\left( o^{\prime },r_{a}\right) <R\left( \varnothing
,r_{a}\right) $ and $R\left( o^{\prime },\pi _{a}\right) >R\left(
\varnothing ,\pi _{a}\right) ,$ 
\begin{eqnarray}
RV\left( y,\left( \pi _{a},p_{-a}\right) \right) &=&RV\left( y,\left(
r_{a},p_{-a}\right) \right) \text{ if }y_{ao}=1,  \label{x1} \\
RV\left( y,\left( \pi _{a},p_{-a}\right) \right) &<&RV\left( y,\left(
r_{a},p_{-a}\right) \right) \text{ if }y_{a\varnothing }=1,  \label{x2} \\
RV\left( y,\left( \pi _{a},p_{-a}\right) \right) &>&RV\left( y,\left(
r_{a},p_{-a}\right) \right) \text{ if }y_{ao^{\prime }}=1.  \label{x3}
\end{eqnarray}

First, we show that $g\left( f^{U}\left( r_{a},p_{-a}\right) ,r\right)
_{a\varnothing }=0$. Suppose not; that is, $y\in \mathcal{Y}^{\ast }\left(
r_{a},p_{-a}\right) $ such that $y_{ao}=1$ for some $o$ satisfying $R\left(
\varnothing ,r_{a}\right) \leq R\left( o,r_{a}\right) $. By Lemma 2, $%
y_{a\varnothing }=1$. Then, by (\ref{x1}), (\ref{x2}) and (\ref{x3}), for
any $y^{\prime }\in \mathcal{Y}^{\ast }\left( \pi _{a},p_{-a}\right) $, $%
y_{a\varnothing }^{\prime }=1$. However, this contradicts that $a$ strictly
prefers $g\left( f^{U}\left( \pi _{a},p_{-a}\right) ,r\right) $ to $g\left(
f^{U}\left( r_{a},p_{-a}\right) ,r\right) $. Therefore, $g\left( f^{U}\left(
r_{a},p_{-a}\right) ,r\right) _{a\varnothing }=0$; that is, for any $y\in 
\mathcal{Y}^{\ast }\left( r_{a},p_{-a}\right) $, $y_{a\varnothing }=0$.

Then, for any $y\in \mathcal{Y}^{\ast }\left( r_{a},p_{-a}\right) $, $%
y_{ao}=1$ implies $R\left( \varnothing ,r_{a}\right) >R\left( o,r_{a}\right) 
$. Since $a$ strictly prefers $g\left( f^{U}\left( \pi _{a},p_{-a}\right)
,r\right) $ to $g\left( f^{U}\left( r_{a},p_{-a}\right) ,r\right) $, for any 
$y^{\prime }\in \mathcal{Y}^{\ast }\left( \pi _{a},p_{-a}\right) $, $%
y_{a\varnothing }^{\prime }=0$ is also satisfied.

Next, we show that for any $y\in \mathcal{Y}^{\ast }\left(
r_{a},p_{-a}\right) $, $y_{ao}=1$ implies $R\left( o,\pi _{a}\right)
<R\left( \varnothing ,\pi _{a}\right) $. Suppose not; that is, $y_{ao}=1$
implies $R\left( o,\pi _{a}\right) >R\left( \varnothing ,\pi _{a}\right) $.
Then, let $y^{\prime }$ be such that $y_{a\varnothing }^{\prime }=1$ and $y_{%
\bar{a}\bar{o}}^{\prime }=y_{\bar{a}\bar{o}}$ for any other $\left( \bar{a},%
\bar{o}\right) .$ By (\ref{x1}), (\ref{x2}) and (\ref{x3}), 
\begin{eqnarray*}
RV\left( y^{\prime },\left( \pi _{a},p_{-a}\right) \right) &<&RV\left(
y,\left( r_{a},p_{-a}\right) \right) \leq \\
RV\left( y^{\prime \prime },\left( r_{a},p_{-a}\right) \right) &\leq
&RV\left( y^{\prime \prime },\left( \pi _{a},p_{-a}\right) \right) ,
\end{eqnarray*}%
for any $y^{\prime \prime }$ such that $y_{a\varnothing }^{\prime \prime }=0$%
, because $y\in \mathcal{Y}^{\ast }\left( r_{a},p_{-a}\right) $. This
contradicts that for any $y^{\prime }\in \mathcal{Y}^{\ast }\left( \pi
_{a},p_{-a}\right) $, $y_{a\varnothing }^{\prime }=0$. Thus, for any $y\in 
\mathcal{Y}^{\ast }\left( r_{a},p_{-a}\right) $, $y_{ao}=1$ implies $R\left(
o,\pi _{a}\right) <R\left( \varnothing ,\pi _{a}\right) $.

We arbitrarily choose $y\in \mathcal{Y}^{\ast }\left( r_{a},p_{-a}\right) $.
By (\ref{x1}), 
\begin{eqnarray*}
RV\left( y,\left( \pi _{a},p_{-a}\right) \right)  &=&RV\left( y,\left(
r_{a},p_{-a}\right) \right) \leq  \\
RV\left( y^{\prime \prime },\left( r_{a},p_{-a}\right) \right)  &\leq
&RV\left( y^{\prime \prime },\left( \pi _{a},p_{-a}\right) \right) ,
\end{eqnarray*}%
for any $y^{\prime \prime }$ such that $y_{a\varnothing }^{\prime \prime }=0$%
. Since $y_{a\varnothing }^{\prime }=0$ for any $y^{\prime }\in \mathcal{Y}%
^{\ast }\left( \pi _{a},p_{-a}\right) $, $\mathcal{Y}^{\ast }\left(
r_{a},p_{-a}\right) =\mathcal{Y}^{\ast }\left( \pi _{a},p_{-a}\right) $ but
this contradicts that $a$ strictly prefers $g\left( f^{U}\left( \pi
_{a},p_{-a}\right) ,r\right) $ to $g\left( f^{U}\left( r_{a},p_{-a}\right)
,r\right) $.

The proof above is satisfied even without the refusal option, because $%
g\left( f\left( \cdot ,\cdot \right) ,r\right) =f\left( \cdot ,\cdot \right) 
$ is satisfied for all $g\left( f\left( \cdot ,\cdot \right) ,r\right) $
above.

\subsection*{Proof of Proposition 7}

Let $f$ be a fair rank-minimizing rule such that, for any $r_{a}\in \mathcal{%
P}$, there is no $p_{a}\in \mathcal{P}$ that dominates $r_{a}$ with the
refusal option. First, we consider the following specific example. Let $%
a_{1},a_{2},a_{3}\in \mathcal{A}$ and $o_{1},o_{2}\in \mathcal{O}$. Suppose $%
q_{o_{1}}=q_{o_{2}}=1$. Let $\delta =\left( o_{1},\varnothing ,\cdots
\right) $ and $\delta ^{\prime }=\left( o_{1},o_{2},\cdots ,\varnothing
\right) $. First, we assume $r_{a_{1}}=p_{a_{2}}=\delta $, $R\left(
o_{2},p_{a_{3}}\right) =1,$ and $R\left( \varnothing ,p_{a}\right) $ for any 
$a\in \mathcal{A\setminus }\left\{ a_{1},a_{2},a_{3}\right\} $ if there
exists some. Then, $\delta ^{\prime }$ is an ODS of $a,$ and 
\begin{eqnarray*}
\left( f\left( \delta ,p_{-a_{1}}\right) \right) _{a_{1}o_{1}} &=&\left(
f\left( \delta ,p_{-a_{1}}\right) \right) _{a_{1}\varnothing }=\frac{1}{2}%
\text{,} \\
\left( f\left( \delta ^{\prime },p_{-a_{1}}\right) \right) _{a_{1}o_{1}} &=&1
\end{eqnarray*}%
Therefore, $a_{1}$ strictly prefers $g\left( f\left( \delta ^{\prime
},p_{-a_{1}}\right) ,r\right) $ to $g\left( f\left( \delta
,p_{-a_{1}}\right) ,r\right) $.

Since $\delta ^{\prime }$ does not dominate $r_{a_{1}}$ under $f$, there is
some $p_{-a_{1}}^{\prime }\in \mathcal{P}^{\left\vert \mathcal{A}\right\vert
-1}$ such that $a_{1}$ strictly prefers $g\left( f\left( \delta
,p_{-a_{1}}^{\prime }\right) ,r\right) $ to $g\left( f\left( \delta ^{\prime
},p_{-a}^{\prime }\right) ,r\right) $. Next, we assume that $%
r_{a_{1}}=\delta ^{\prime }$. Then, $\delta $ is an OPS of $a_{1}$.
Therefore, $a_{1}$ strictly prefers $g\left( f\left( \pi
_{a},p_{-a_{1}}^{\prime }\right) ,r\right) $ to $g\left( f\left(
r_{a_{1}},p_{-a}^{\prime }\right) ,r\right) $, where $r_{a_{1}}=\delta
^{\prime }$ and $\pi _{a}=\delta $.

The proof above is satisfied even without the refusal option, because $%
g\left( f\left( \cdot ,\cdot \right) ,r\right) =f\left( \cdot ,\cdot \right) 
$ is satisfied for all $g\left( f\left( \cdot ,\cdot \right) ,r\right) $
above.

\section*{Declaration of generative AI and AI-assisted technologies in the writing process}

During the preparation of this work the author used ChatGPT (OpenAI) in order to improve readability and language. After using this, the author reviewed and edited the content as needed and takes full responsibility for the content of the publication.

\end{document}